\documentclass[12pt,twoside,english]{iopart}
\usepackage{iopams}
\usepackage{mathptmx}

\usepackage[latin9]{inputenc}
\usepackage{units}
\usepackage{amsthm}

\usepackage{amssymb}
\usepackage{graphicx}
\usepackage{caption}
\usepackage{color}
\usepackage[position=t,singlelinecheck=off]{subfig}

\makeatletter
\renewcommand{\fnum@figure}{g. \arabic{figure}}

\@ifundefined{textcolor}{}
{%
 \definecolor{BLACK}{gray}{0}
 \definecolor{WHITE}{gray}{1}
 \definecolor{RED}{rgb}{1,0,0}
 \definecolor{GREEN}{rgb}{0,1,0}
 \definecolor{BLUE}{rgb}{0,0,1}
 \definecolor{CYAN}{cmyk}{1,0,0,0}
 \definecolor{MAGENTA}{cmyk}{0,1,0,0}
 \definecolor{YELLOW}{cmyk}{0,0,1,0}
}

\@ifundefined{date}{}%{\date{}}
\makeatother

\usepackage{babel}
\begin{document}

\title[Heterogeneous network games with conflicting preferences]{Equilibria, information and frustration in heterogeneous network games with conflicting preferences}

\author{M Mazzoli\textsuperscript{1} and A S\'anchez\textsuperscript{2,3}\\ \ \\ }

\address{$^1$Universit\'a degli Studi di Torino, Dipartimento di Fisica, I-10125 Torino, Italy\\ \ \\
$^2$Grupo Interdisciplinar de Sistemas Complejos (GISC), Departamento de Matem\'aticas and Institute UC3M-BS for Financial Big Data, Universidad Carlos III de Madrid, Legan\'es, 28911 Madrid, Spain\\ \ \\
$^3$Instituto de Biocomputaci\'on y F\'\i sica de Sistemas Complejos (BIFI), Campus R\'\i o Ebro, Universidad de Zaragoza, 500018 Zaragoza, Spain}

\vspace{10pt}
\begin{indented}
\item[]June 2017
\end{indented}

\begin{abstract}
Interactions between people are the basis on which the structure of our society
arises as a complex system and, at the same time, are the starting point of any physical 
description of it. In the last few years, much theoretical research has addressed
this issue by combining the physics of complex networks with a description of interactions in
terms of evolutionary game theory. We here take this research a step further by introducing a
most salient societal factor such as the individuals' preferences, a characteristic
that is key to understand much of the social phenomenology these days. We consider a
heterogeneous, agent-based model in which agents interact strategically with their neighbors 
but their preferences and payoffs for the possible actions differ. We study how such a heterogeneous
network behaves under evolutionary dynamics and different strategic
interactions, namely coordination games and best shot games. With this model we study
the emergence of the equilibria predicted analytically in random graphs under best response
dynamics, and we extend this test to unexplored contexts like proportional imitation and scale
free networks. We show that some theoretically predicted equilibria do not arise in simulations
with incomplete Information, and we demonstrate the importance of the graph topology and
the payoff function parameters for some games. Finally, we discuss our results with available
experimental evidence on coordination games, showing that our model agrees better with the 
experiment that standard economic theories, and draw hints as to how to maximize social
efficiency in situations of conflicting preferences.
 \end{abstract}

\vspace{2pc}
\noindent{\it Keywords}: agents, agent, coordination, proportional imitation, best response, network, scale free
%\keywords{agents, agent, coordination, proportional imitation, best response, network, scale free}
%\maketitle

\section{Introduction}

The behavior of complex systems is determined by its components and, chiefly, by their interactions. Generally speaking, specifying the interactions of a complex system \cite{mitchell:2011} involves a network, that indicates who interacts with whom, and the rule or law governing the interaction itself. This paradigm applies to purely physical systems but also to social systems \cite{miller:2007,castellano:2009}, the difference being that in the latter case interactions are strategic, i.e., the agents have some degree of intelligence and can anticipate the reactions of their counterparts to their own actions. Such a situation requires a description in terms of game theory \cite{shecter:2016} and, in fact, this framework it is becoming the standard to describe complex systems in social and economic systems \cite{vega-redondo:2007,goyal:2007,jackson:2008,easley:2010}.  

The complex systems community has devoted a lot effort to this approach in this century (see, e.g., \cite{szabo:2007,roca:2009,perc:2013,perc:2017} for reviews). Typically, the models considered in this research are a combination of the above mentioned ingredients of games (describing how interactions take place) and networks (describing the interaction structure) with some evolutionary dynamics \cite{nowak:2006}. The rationale for such an approach is twofold: On one hand, several of the dynamics can be shown to lead to equilibrium states that are related to the Nash equilibria of the network game \cite{shecter:2016,nash:1950}, i.e., to what the system should be actually doing were it formed by rational agents. On the other hand, a dynamical approach is intended to explain which, if any, such equilibria are actually reached by pointing to a mechanism that shows how they can be reached by agents whose cognitive capabilities are bounded, i.e., they do not conform to the omniscient rational agents of economics. 

This type of approach is currently being applied to understand different socially relevant issues, such as the emergence of cooperation \cite{axelrod:1984}, where a paper on spatiotemporal chaos \cite{nowak:1992} originated a huge number of papers on theoretical models \cite{szabo:2007,roca:2009}. This effort further fructified in several experiments with human subjects \cite{grujic:2010,gracia-lazaro:2012,grujic:2014} leading to the understanding of the dynamics in terms of moody conditional cooperation and reinforcement learning \cite{cimini:2014,ezaki:2016,horita:2017}. In this context, a very pressing issue that is key to understand human societies and how they can be nudged towards cooperating with each other is that of identity (religious, linguistic, political, etc.) as the source or reason for different preferences \cite{akerlof:2000}. 
Indeed, the interplay between our preferences and the influence of our social relationships (friends, acquaintances, coworkers) on our choices arises in may aspects of our daily life life. This occurs, for instance, when we choose friends \cite{mcpherson:2001} or neighbors
\cite{schelling:1978}, a process where individual preferences are a key in our decisions. 
Another example of the importance of preference is the large influence peers have on human behavior 
\cite{jackson:2008}, affecting whether people's behavior aligns to that of their social relationships
\cite{morris:2000}. Such social influence effects range from which products we buy \cite{key-3}, to the decision to get involved in criminal activities \cite{ballester:2006}, or to
our participation in collective action \cite{granovetter:1978}. Particularly important is the case of strategic interaction in networks, a realization of which is the situation in which one has to decide on  a technological product
that should be compatible with the co-workers's choices. This is a case of a coordination problem, in the class we will discuss below, and clearly choices change depending on
others' decisions, but every person has her own initial preference \cite{vives:2009}. All these particular situations boil down to a specific research question: What is the effect of individual preferences on strategic interaction, be it of the coordination or anti-coordination type?

We here study this issue in a very broad range of socially relevant scenarios, by using the model introduced by Hern\'andez {\em et al.} \cite{key-2}. This is a generalization of earlier work \cite{key-3} where two entire classes of games were studied, namely coordination games and social dilemmas (more precisely, strategic complements and strategic substitutes) in random networks. In \cite{key-2}, the problem of diversity in preferences was analyzed by considering that there are two different types of players in the population, and that each type prefers (because the corresponding payoff is larger) one of the two available actions. Therefore, coordination and/or cooperation becomes more difficult, in so far that agents have incentives to choose a specific action that yields more benefit to them irrespective of the choices of those with whom they interact. In fact, as has been recently shown \cite{hernandez:2017}, this difficulty in coordination is predicted to be largely dependent on the payoff ratio between the preferred and the disliked actions, but it may even disappear when payoffs become similar. In this context, we here address a number of issues that are relevant from viewpoints of both the evolutionary dynamics of complex systems and its application to societal issues. Firstly, we intend to identify the effect of the presence of agents with different preferences in the system and how this effect depends on the network structure. Secondly, we want to understand whether these effects change in cases where all agents would prefer to choose the same action as the rest (complements/coordination) or different actions (substitutes/anti-coordination). Last, but not least, we want to assess the relevance and validity of our approach by comparing with existing experimental results.

Our results are organized as follows: we begin by introducing the main game theory, focusing on the conceptual
differences the preferences paradigm brings in to the homogeneous framework (Sec.\ II). We then study in Sec.\ III the dynamically relevant equilibria
and compare them with the ones found in  \cite{key-2,hernandez:2017} with an analytical, static approach. Subsequently, we also extend
the study to the case of a scale-free network and Proportion Imitation, comparing the differences when the same games are played in the absence of individual preferences. 
Our next step is to look into the case of incomplete information (Sec. IV), where agents do not have knowledge about their neighborhood, which we analyze having the complete information situation case as our reference point. 
% {\color{red} We stress, as will be discussed below, that we mean information in the sense that the knowledge agents have on the others' behavior and payoffs, not in the sense on Shannon.}  
In this case, only best response case because, as we will discuss below, proportional imitation cannot be applied for lack of information. In the conclusion (Sec.\ V) we summarize our results, compare with the available experimental results, and discuss how they give insight on how to solve coordination problems in situations of conflicting preferences.

\section{Model}

The building blocks of our models are a set of agents, a game that specifies their interaction, and a network of connections between them that rules who interacts with whom. Each agent has two possible actions, which we label $X=\{0,1\}$ and a preference for one of them. Due to this preferential heterogeneity, individuals who choose their preferred action gain greater payoffs than when they choose the other one, for every game in the families we will consider below. This is mathematically represented by two parameters, which represent the rewards for choosing the
liked or disliked option, and hence affect the incentive to change action and so the dynamics of the game.
In order to cover an ample set of games (i.e., of possible interactions between people) we work with a very general payoff function
\begin{eqnarray} 
\fl u_{i}(\theta_{i},x_{i},x_{N_{i}})=\lambda_{x_{i}}^{\theta_{i}}[1+\delta\sum_{j\in k_{i}}I_{\{x_{j}=x_{i}\}}+(1-\delta)\sum_{j\in k_{i}}I_{\{x_{j}\neq x_{i}\}}] 
\label{eq1}
\end{eqnarray}
where $x{}_{i}$ is the action taken by  agent i, $k{}_{i}$ are
the neighbors of agent $i$ as specified by the corresponding network, $x{}_{N_{i}}$ is the vector of actions
taken by $i$'s neighbors, $\theta_{i}$ is agent $i$'s preference (that, as actions, can be 0 or 1), and $I_{\{x_{j}=x_{i}\}}$indicates
the neighbors who choose the same action as $i$. As for the payoffs $\lambda$ takes
the value  $\alpha$ if the agent takes his liked action or $\beta$
otherwise, where $0<\beta<\alpha<2\beta$, and $\delta$ defines the kind
of game we are playing: if $\delta=1$ we are playing a Coordination
Game (CG; the best action is to do as others do), if $\delta=0$ we are playing Anticoordination Game (AG; the best action is to do the opposite of what the others do). We note that in economics jargon these game families are usually referred to as strategic complements and strategic substitutes, respectively \cite{key-3}, but in this work we prefer to use the names above as they make it easier for the reader to grasp the actual meaning of the two types of interaction.  We also note that the original homogeneous model in \cite{key-3} is recovered when all players have the same preference.  
Below, we will in addition differentiate two types of arrangements: The first is a situation
of complete information, where every individual knows 
who his neighbours are and what they do at every round of the game; the second is a situation
of incomplete information, where agents know how many neighbors do they have and the distribution of preferences in the network,
but they don't know what the specific preferences of their neighbors are. In this last case, agents
can infer the proportion of neighbors who might prefer 
action 1 or 0 knowing the preferences distribution and their degree, but they don't know exactly the distribution
of actions in their local network. 
Finally, we need to specify the dynamics we will consider in this model. Following \cite{us:2014}, we will use the following two dynamics as representative of the more economics-style (best response) and evolutionary (imitation) choices.

\subsection{Best response}

Let us call $\chi_{i}$ the number of agent $i$'s neighbors who choose action
1, so the number of neighbors that choose action 0 is $k{}_{i}-\chi_{i}$.
As described in \cite{key-2}, from the purely static, theoretical viewpoint in economics, we have two thresholds to compare with $\chi_{i}$
in order to permit to agent $i$ to decide which action to take:

\begin{eqnarray} 
\underline{\tau}(k_{i})&=&\left[\frac{\beta}{\alpha+\beta}k_{i}-\frac{\alpha-\beta}{\alpha+\beta}\right],  %\tag{2}\
\label{2}\\
\overline{\tau}(k_{i})&=&\left[\frac{\alpha}{\alpha+\beta}k_{i}+\frac{\alpha-\beta}{\alpha+\beta}\right]. %\tag{3}
\label{3}
\end{eqnarray}
\\
With these two thresholds, the best response for agent i with preference $\theta_{i}=1$ in a CG is given
by:

\begin{eqnarray}
%\tag{4}
\label{4}
x_{i}= \left\{\begin{array}{@{}l@{\quad}l}
1 \quad \mbox { iff$\;\chi\geq\underline{\tau}(k_{i})$}\\  
0 \quad \mbox {otherwise; }
\end{array} \right.
\end{eqnarray}
\\
Conversely, the best response for agent i with preference $\theta_{i}=0$ is given
by:

\begin{eqnarray}
%\tag{5}
\label{5}
x_{i}=\left\{\begin{array}{@{}l@{\quad}l}
0 \quad \mbox{ iff$\;\chi_{i}\leq\overline{\tau}(k_{i})$}\\
1 \quad \mbox {otherwise.}
\end{array}\right.
\end{eqnarray}
\\
These two options for the CG are simply illustrated in the sketch of figure 1.
below.
\begin{figure}[h]
\centering
\includegraphics[width=5in]{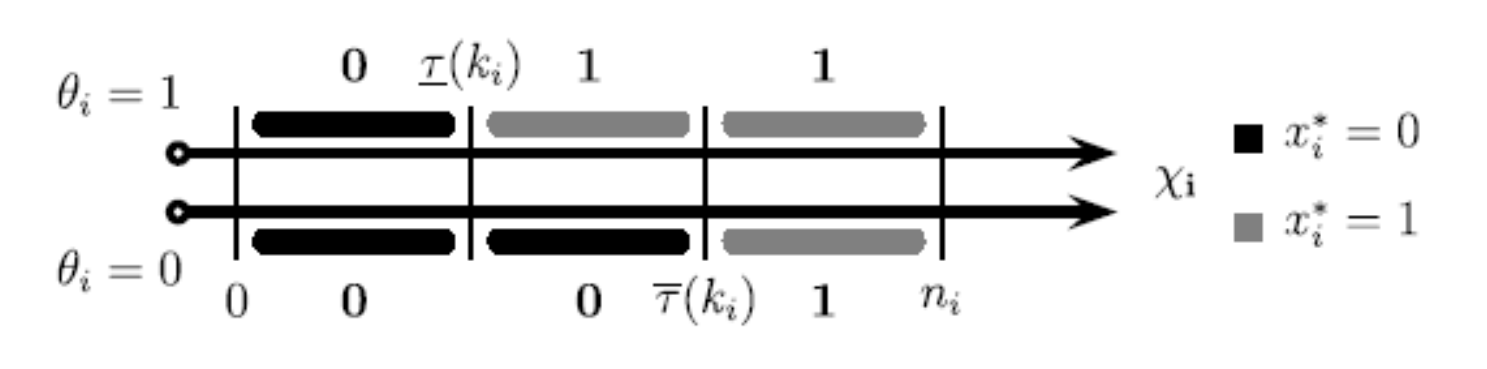}
\caption{Action change thresholds for both preferences in the Coordination Game.}
\end{figure}

Similarly, in the case of an AG,  agent $i$ will choose the liked
action when $\chi_{i}\leq\overline{\tau}(k_{i})$ for $\theta_{i}=1$
and when $\chi_{i}\geq\underline{\tau}(k_{i})$ for $\theta_{i}=0$. With these results, the predictions from the analysis in \cite{key-2} are that 
equilibria in the network are such that all players coordinate
on one action (specialized) or both actions are chosen by different
players (hybrid). There are two categories of equilibria, depending
on whether all players coordinate in choosing the action they like
(satisfactory) or at least one player chooses the disliked option
(frustrated). So we have four possible equilibria: (i) satisfactory
specialized (SS) where all players coordinate on the same action,
which is their preferred choice (so, this can happen only in the homogeneous model, where all agents have the same preference); (ii) frustrated specialized (FS),
where all players coordinate on the same action, but at least one
of them is choosing his disliked option; (iii) satisfactory hybrid
(SH), where all players choose their preferred option but there is
at least one player with a preference different from the rest, so that
both actions are present; and (iv) frustrated hybrid (FH) which presents
both actions and at least one player chooses her disliked option. We will analyze below what happens dynamically, i.e., when the game is repeated for a number of rounds starting from a random initial condition and players choose their action in their next round through myopic best response \cite{roca:2009b}, by deciding their next action as a best response to their neighbors' actions in the previous round. 

\subsection{Proportional imitation}

The second dynamics we will consider in this study consists of the imitation
of a neighbor: at each time step a fraction of the  agents choose one of their neighbors at random
and, if the neighbor's payoff is higher than her payoff, she chooses
the neighbor's action for the next time step with a probability given
by the difference between their two payoffs, according to 

\begin{eqnarray}
\label{6}
P\{\pi\rightarrow\pi_{i}^{(t+1)}\}= \left\{\begin{array}{@{}l@{\quad}l}
(\pi_{j}^{(t)}-\pi_{i}^{(t)})/\phi  & \mbox{ if }\pi_{j}^{(t)}>\pi_{i}^{(t)},\\
0 & \mbox{otherwise.}
\end{array}\right.
\end{eqnarray}

The reason to consider this dynamics is that it is the evolutionary version of the well-known replicator dynamics \cite{shecter:2016} that, in the limit of an infinite number of agents, can be shown to converge to the Nash equilibria of the game. However, the approach to the equilibria is different from the best response case: In best-response, all agents try to choose directly the action that would give them the largest payoff given the actions of the others, whereas in imitation dynamics agents have a much smaller cognitive capability and limit themselves to imitate some action that they perceive to yield higher payoffs. Imitation is thus a much more realistic dynamics to represent human (or even animal or bacteria) decisions as arising from something akin to a learning (or adaptation) process. On the other hand, best response is deterministic whereas imitation dynamics is stochastic, which provides another interesting comparison.

\subsection{Simulations}

In what follows, we report the results of a simulation program in which we have looked at the behavior of the model for its most important parameters: 
the payoff
to choose the liked option, $\alpha$, the payoff to choose the disliked option, $\beta$,
and the proportion of 1-preference agents $\rho$, always respecting the conditions of the games: $0<\beta<\alpha<2b\; and\;1<\rho<0$. We have considered networks with $n=10^{2}$ nodes (except when specified)
and all the results are averaged over a number of iterations of $t=10^{2}$
time steps each. Simulations with more nodes have also been done and will be reported below in order to check our conclusions, and the number of iterations has been taken to be 20 or 50, also for verification purposes. Simulations are run over 8$\times$ 8 different
sets of values for $\alpha$ and $\beta$, choosing $0.2\leq\alpha\leq0.9$ and
$\nicefrac{\alpha}{2}<\beta<\alpha$, and they are as long as needed for the system to equilibrate. We verified our code by checking that, for the homogeneous model, we recovered the results reported in \cite{us:2014}, with very satisfactory results. In addition, we considered two types of networks: an Erd\"os-Re\'nyi (ER) \cite{Erdos1960} random graph, for different values of connectivity $m$, and  a Barab\'asi-Albert (BA) \cite{Barabasi1999} scale free graph, with 3 (the parameter typically referred to as $m$ in the BA model, different from the connectivity $m$ of our ER networks) edges connecting every new node added to the graph to nodes already in the network. The reason to do this is that the theoretical predictions summarized above are only valid for an uncorrelated random graph, which is represented by the ER network. Therefore, we find it interesting to include a completely different network such as the BA one, that is in fact associated to many more realistic social situations. 

\section{Complete information}

We begin by discussing the results when players have complete information about their neighbors's preferences. For the sake of clarity, we present the results separately for each type of game and each type of dynamics. In this section and throughout the paper, 
in all plots
every dot is a particular set of parameters. In equilibria graphics,  red dots are the Specialized Satisfactory equilibria,  yellow
ones are the Specialized Frustrated equilibria, green ones are
the Hybrid Frustrated equilibria, and blue ones are potentially Hybrid
Satisfactory equilibria or Hybrid Frustrated. 

\subsection{Coordination Game}

\subsubsection{Best Response}

\begin{figure}[h]
{\centering
(a)\subfloat{\includegraphics[width=3in]{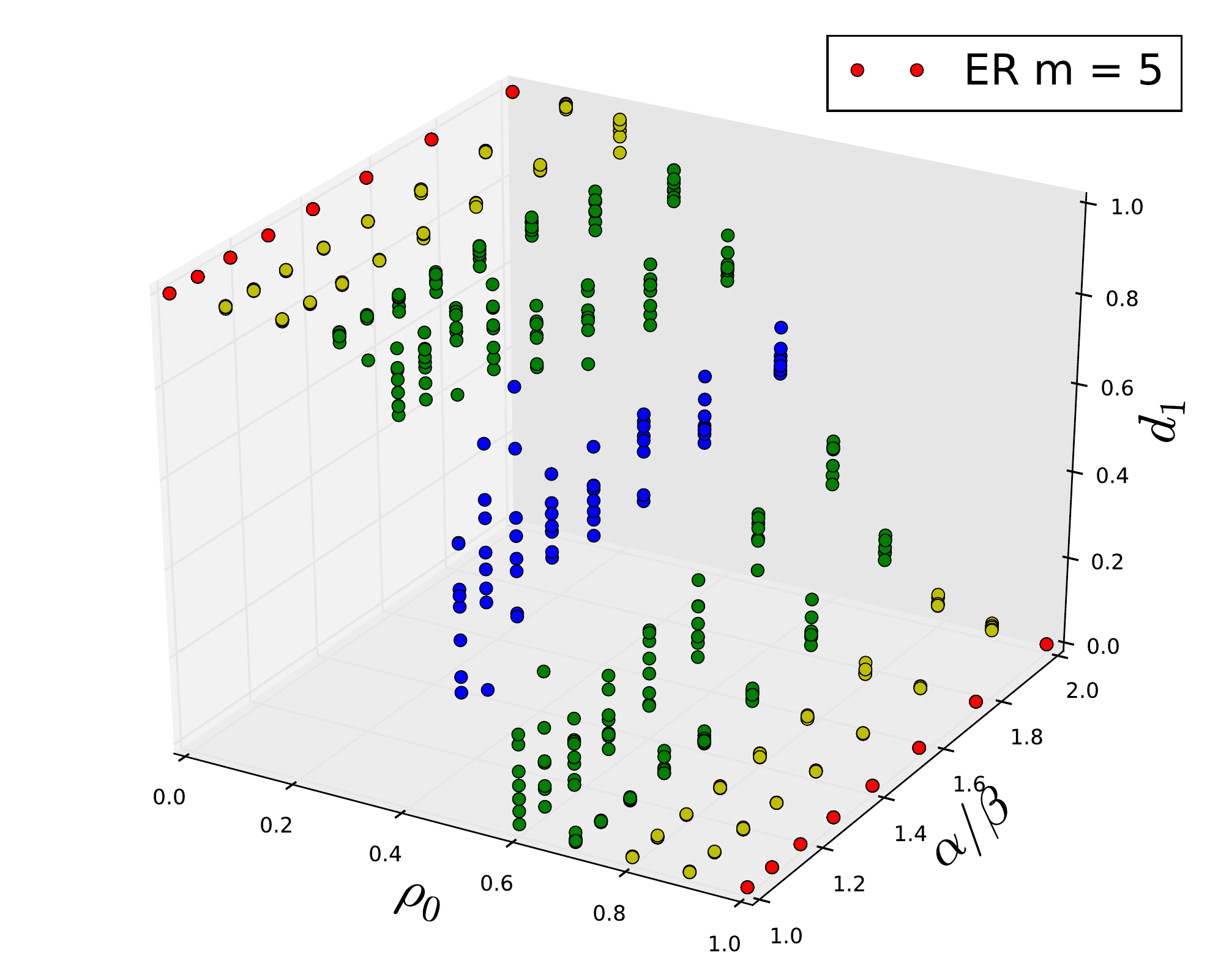}}
(b)\subfloat{\includegraphics[width=3in]{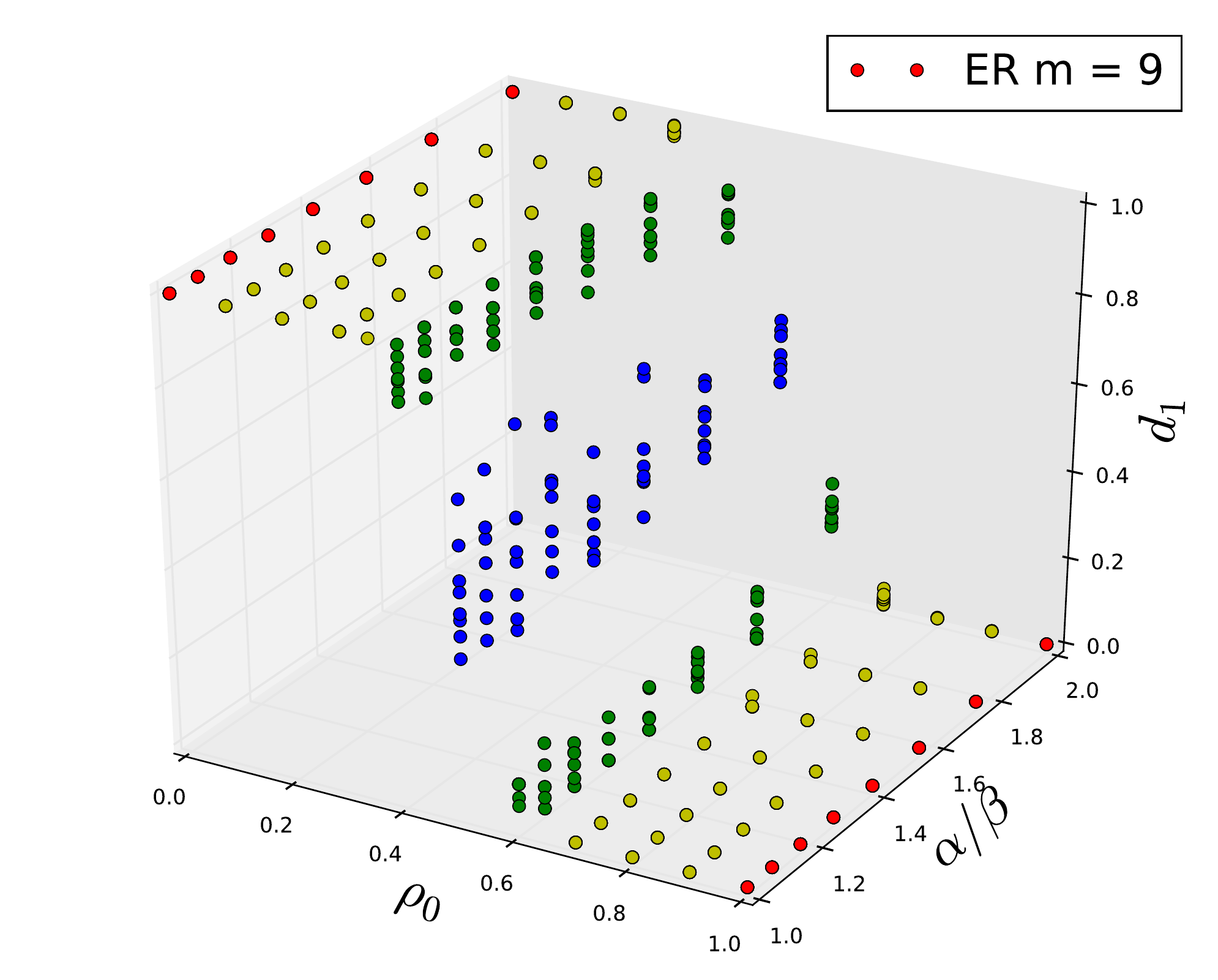}}\\
(c)\subfloat{\includegraphics[width=3in]{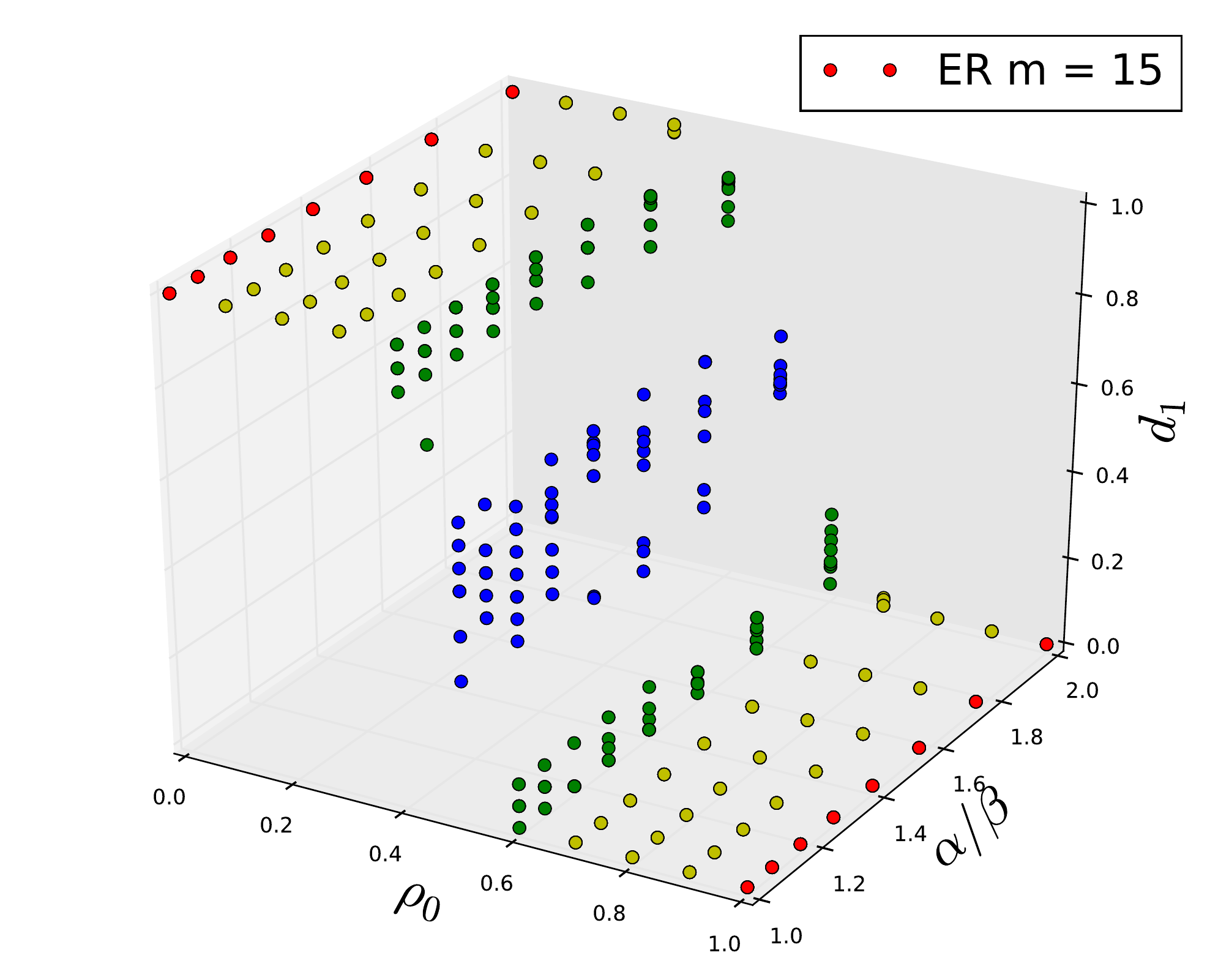}}
(d)\subfloat{\includegraphics[width=3in]{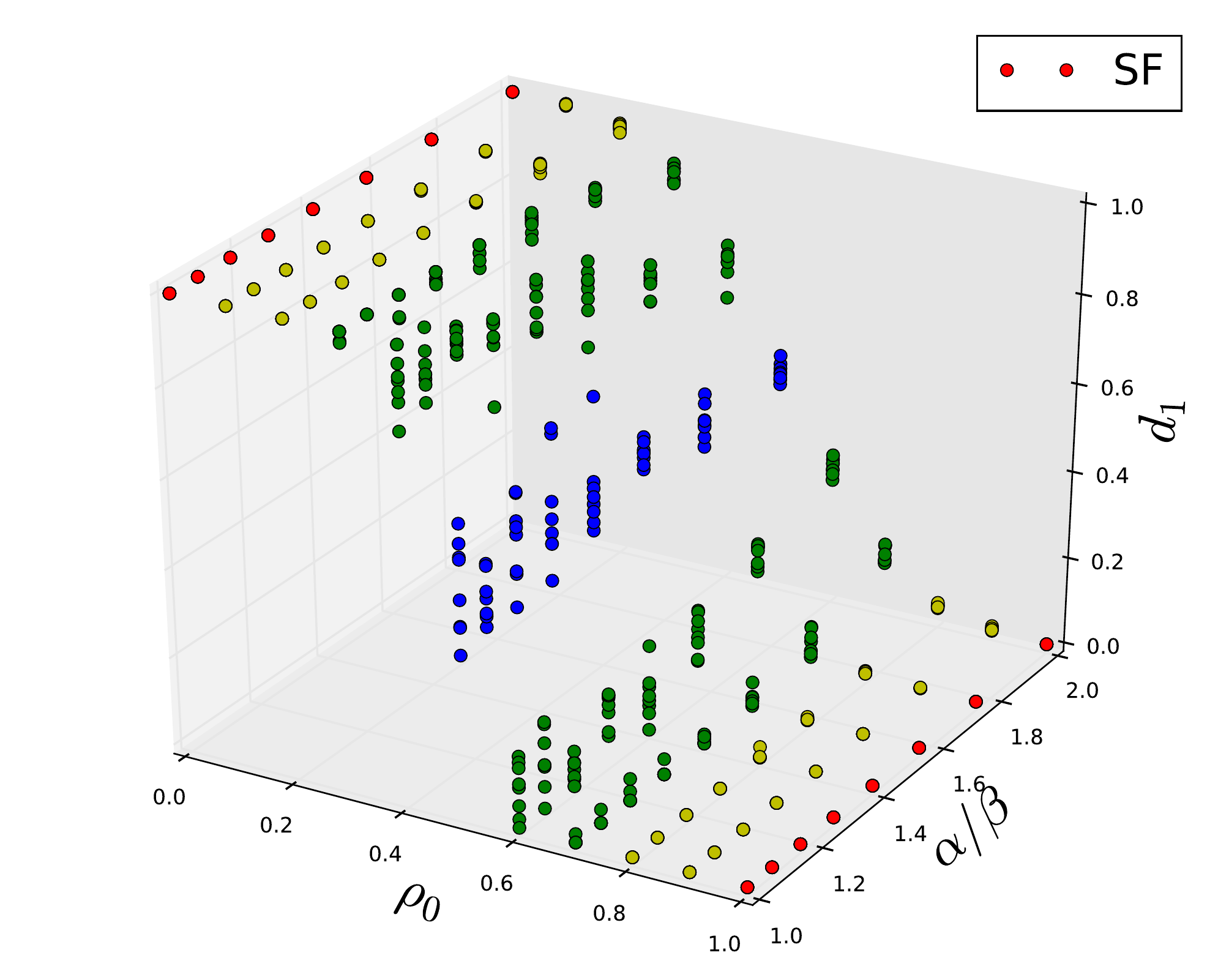}}
}
\caption{Final average density of agents who play action 1 $d_1$ against fraction of 0-preference players $\rho_0$ and reward ratio $\alpha/\beta$ in equilibrium, for the CG on different ER networks (connectivity as indicated in the plot) and a BA network. Red: Specialized Satisfactory equilibria;  yellow:
Specialized Frustrated equilibria; green, Hybrid Frustrated equilibria; blue, potentially Hybrid
Satisfactory equilibria or Hybrid Frustrated.}
\end{figure}
In order to compare with the predicted results from the static approach, we discuss first uncorrelated networks given by the ER model. Our first observation is that, 
as we raise the connectivity of  ER networks,  equilibria tend to be more specialized and  hybrid
equilibria tend to disappear. Figure 2 shows clearly that  equilibria
are symmetric for different fractions of preferences,  as it was to be expected as there is nothing intrinsically different between the two types. When the two preferences have an equal number of agents in the population, the final
density of agents who choose action 1 in equilibrium takes values in a range around 0.5, a range that tends to decrease the more we raise the ratio $\alpha/\beta.$ This is due to the particularities of the different networks realized in the simulations, as there may be local environments that just by chance make agents choose one action that is not their preferred one. 

It is important to realize that 
the ratio $\alpha/\beta$ gives us a direct measure of the incentive given to agents
to maintain their preferred action instead of changing it, so it is clear that when this incentive is small a larger variety of outcomes are possible. For instance, when $\alpha/\beta\rightarrow1$,
if the simulation starts with a 60\% of 0 agents, the final equilibria
will be a 0-specialized frustrated one, because the 40\% of initial
1 agents aren't sufficiently motivated to maintain their liked
option. On the contrary, when $\alpha/\beta\rightarrow2$, in the
same case of a 60\% of initial 0 preferences, a relevant part of the
1 agents resist the temptation
to go against their preferences, because the incentive to maintain their preferred action
is really higher than the one given to change (unless in very specific realizations one 1 agent is surrounded by a large number of 0 agents; this is something that occasionally, but not frequently, will occur). Therefore,  the final equilibrium
is not specialized anymore, but it is hybrid and there will be less
frustration in the final state. This is seen in figure 2 by the fact that for small  $\alpha/\beta$ the transition from one specialized equilibrium in the action of the majority of the agents to the other is much more abrupt than for large $\alpha/\beta$, implying that the range of fractions of each type leading to hybrid equilibria is larger in the latter case. This is so 
specially in the less connected graphs: in the case of $\alpha/\beta\rightarrow2$
in the final states there are more satisfied agents, because they
are pushed to maintain their action as they do not have many neighbors which could induce them to change. We can say
the opposite in the case with $\alpha/\beta\rightarrow1$ where in
the final states we find more frustrated agents and specialized
equilibria. Figure 3 confirms this insight by showing the equilibria for the two extreme values of the payoff ration and including the density of frustrated agents. For small $\alpha/\beta$, the fraction of frustrated agents grows approximately linearly with the fraction of 0 agents until they reach a 50-50 distribution: this makes sense as for small payoff ratios only a small majority of agents of the opposite preference is needed to make one change. Interestingly, frustration is much lower for larger $\alpha/\beta$, reflecting the fact that locally it may pay to keep one's preferred action even if there are more neighbors of the opposite side. This is particularly true for low connectivity networks; larger connectivity leads to larger chance to have many neighbors of the opposite type forcing one to change (keep in mind that best response is deterministic and always chooses what is best in view of the environment). 

Let us now look at the case of the BA scale free graph. As we can see from the plot, the overall behavior
is not far from the ER random graph one with $m = 5$. This is likely to result from the fact that there are a large majority of agents that have a small number of neighbors, and therefore in terms of the total fraction of agents choosing each action these subset of nodes dominates the dynamics. On the contrary, hubs are just one neighbor of other agents, so in a best response environment their contribution to the decision of their neighbors is not particularly relevant. In this manner, we have identified two main variables that determine the type of
equilibria that will come out in a Coordination Game: the connectivity
of the graph and the payoff ratio. 
\begin{figure}[h]
{\centering
(a)\subfloat{\includegraphics[width=3in]{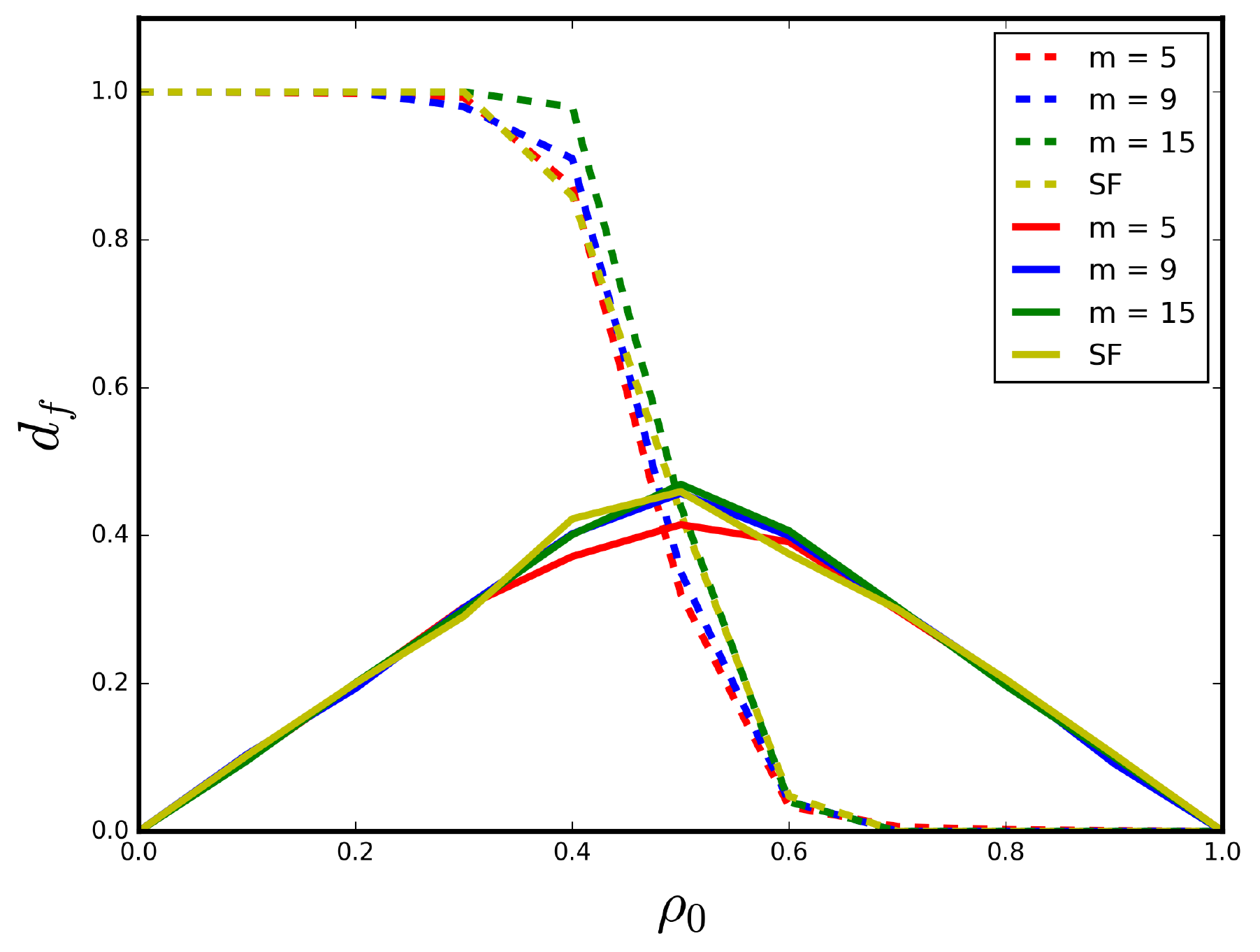}}
(b)\subfloat{\includegraphics[width=3in]{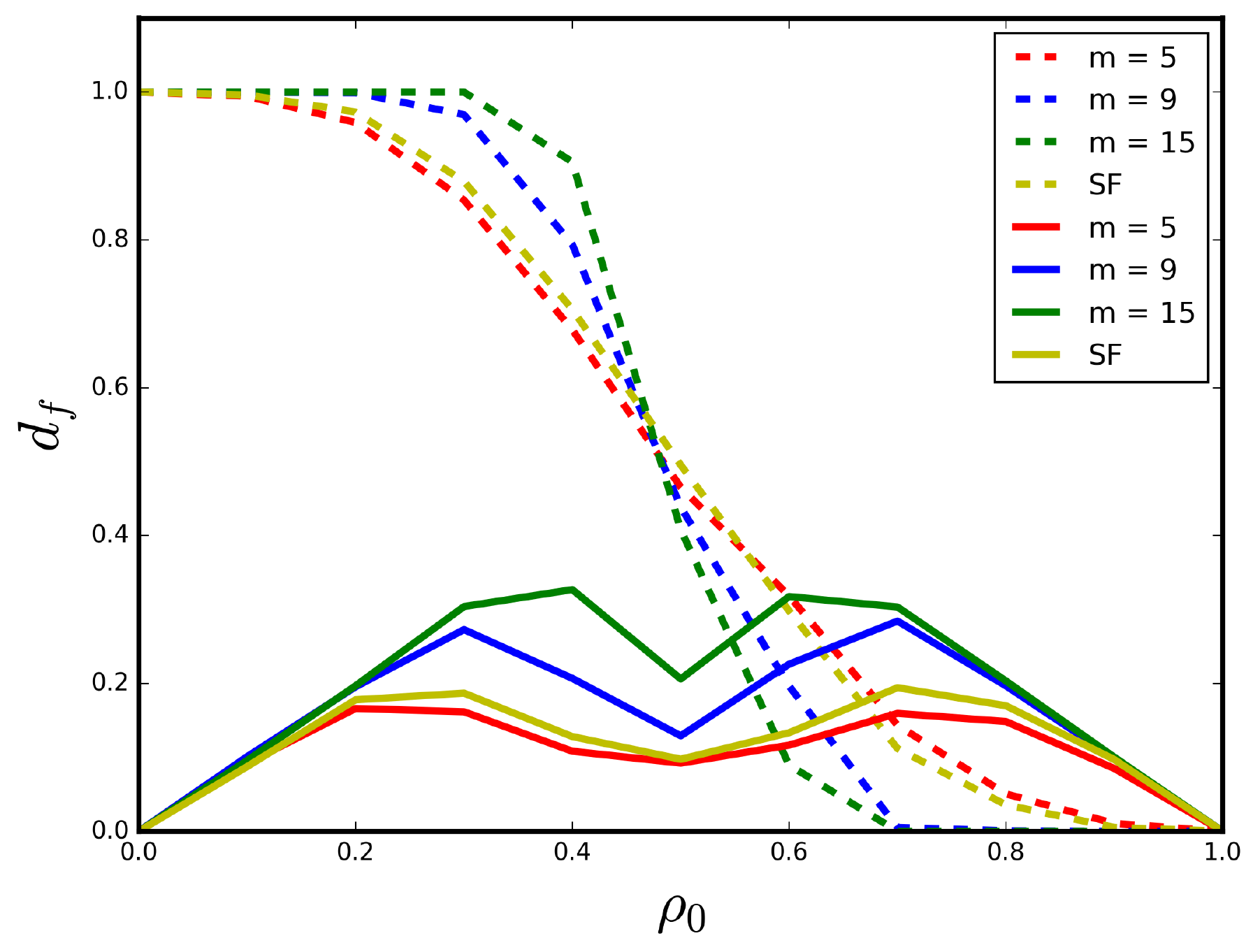}}\\
}
\caption{Final average density of frustrated agents $d_f$ over
50 realizations against 0-preference density $\rho_0$ (solid lines). Shown also is the corresponding final average density $d_1$ (dashed lines). (a) Coordination Game with reward ratio $\alpha/\beta=1$. (b) Coordination Game with reward ratio $\alpha/\beta=2$. Lines are as indicated in the plot.}
\end{figure}

In the light of what we know about the homogeneous model \cite{key-2}, we observe that in both models connectivity
is a catalyst for the achievement of a specialized equilibrium: the more the connectivity is, the less the hybrid
equilibria will be. What was true in the homogeneous case about cooperation (understood as coordination in the Pareto-dominant or more profitable equlibrium) can be also said of the heterogeneous
model about coordination: If full cooperation was reached thanks
to high connectivity under the same cooperation incentive, now high
connectivity allows, under the same reward ratio, to reach full coordination
(which means specialized equilibria) in the most of the cases.
On the other hand, an important difference with the previous
model is that in the homogeneous case agents had an incentive to cooperate ($\alpha$)
which helped the achievement of full cooperative final states. When preference enters the game,
there is a payoff ratio which hinders full
coordination, because it preserves the satisfaction of the individual. Therefore, preference does qualitatively change the problem and, more importantly, the perception of the outcome of evolution as satisfactory by the individuals.

%\subsubsection*{\centerline{(a). Analogies and differences with the homogeneous model}}

%\subsubsection*{\centerline{(b). Raising the size of the network to 1000 nodes}}

In order to verify the above conclusions, wee raised the number of agents to $n = 10^3$ to see if network size could affect the final equilibria.
\begin{figure}[h]
{\centering
(a)\subfloat{\includegraphics[width=3in]{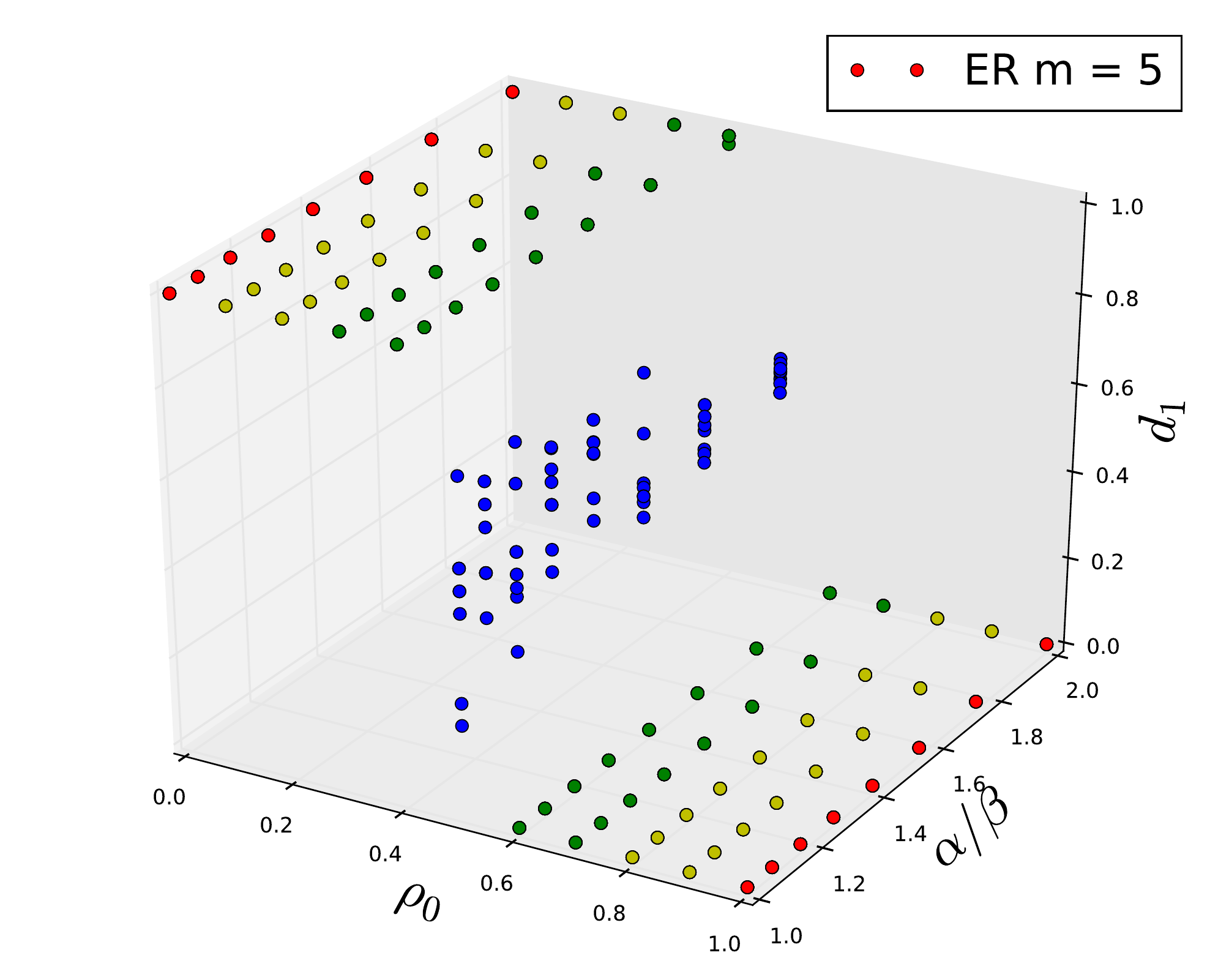}}
(b)\subfloat{\includegraphics[width=3in]{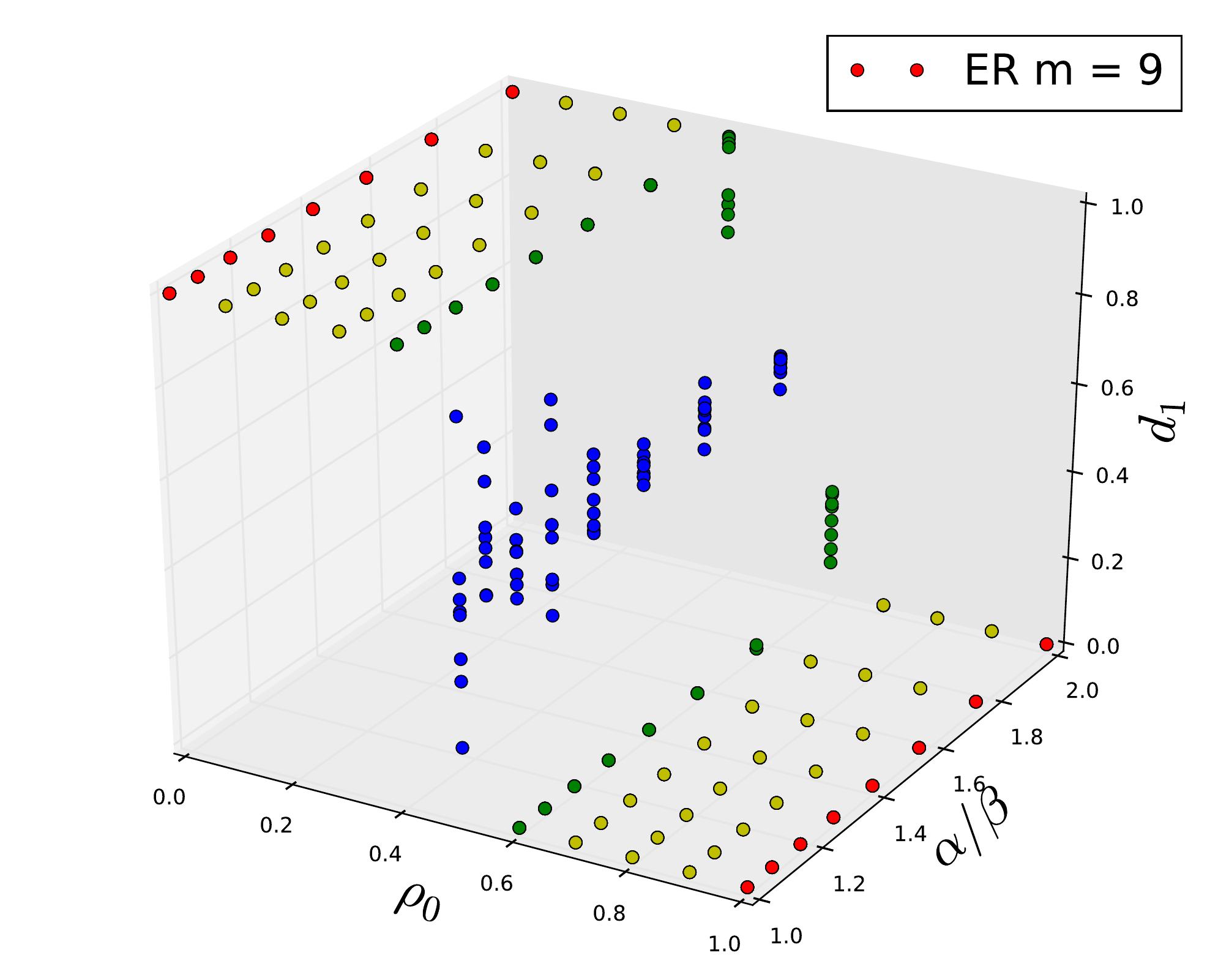}}\\
(c)\subfloat{\includegraphics[width=3in]{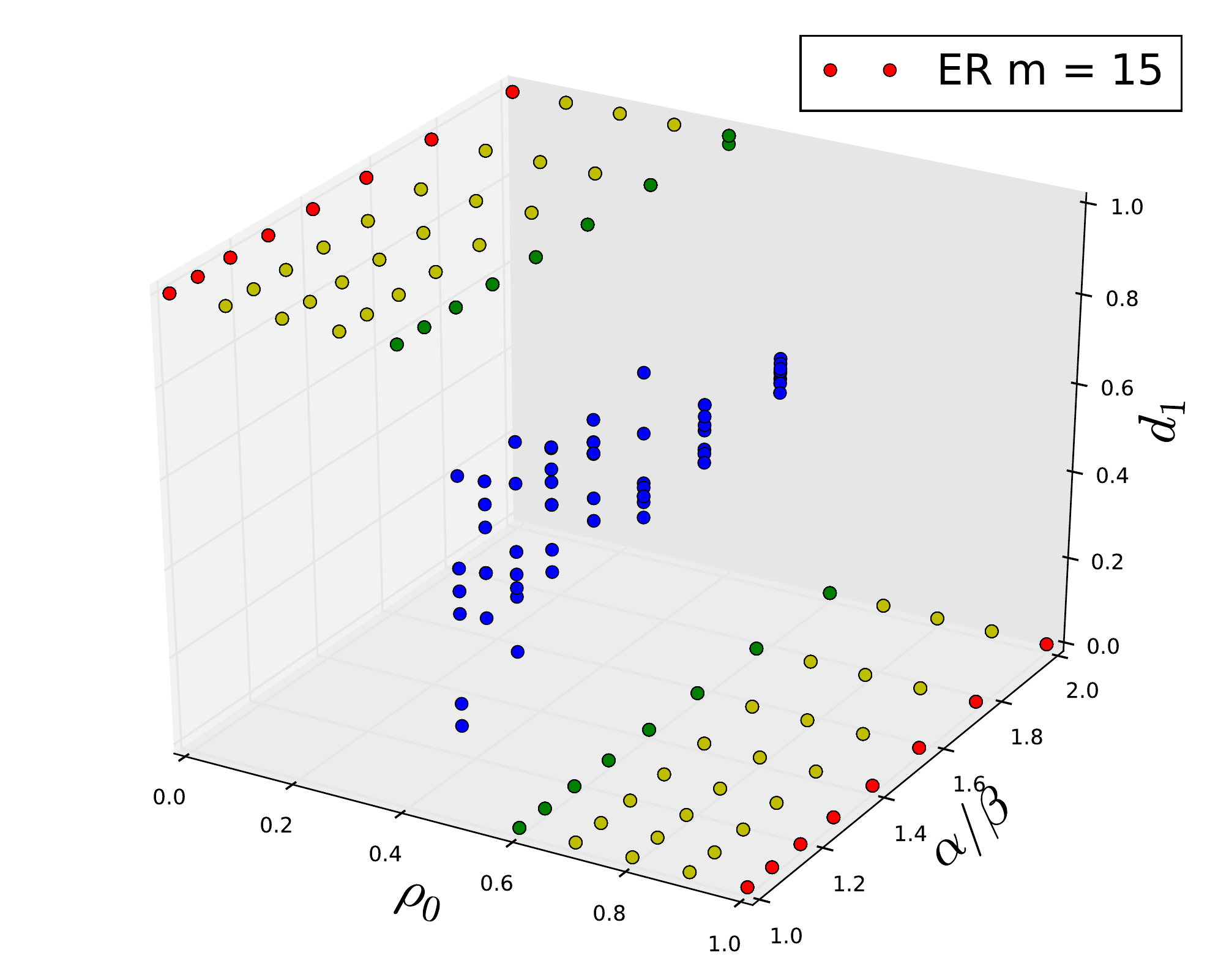}}
(d)\subfloat{\includegraphics[width=3in]{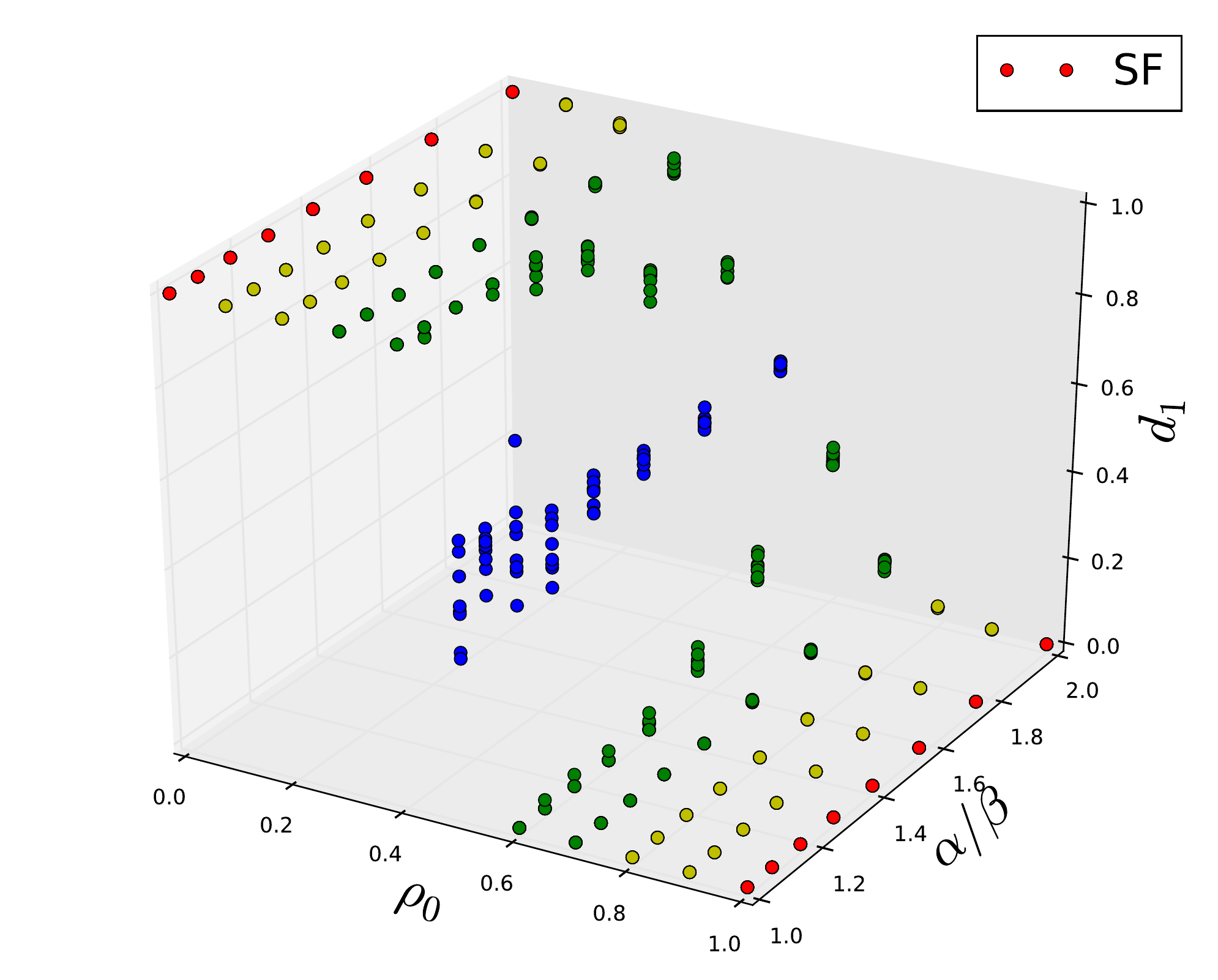}}
}
\caption{Final average density of agents who play action 1 $d_1$ against fraction of 0-preference players $\rho_0$ and reward ratio $\alpha/\beta$ in equilibrium, for the CG on different ER networks (connectivity as indicated in the plot) and a BA network. Simulations with 1000 agents. Colors as in figure 2.}
\end{figure}
Comparing graphics in figure 4 with those with $n = 10^2$ nodes we can see
that the size of the network fosters coordination. Hybrid equilibria almost disappear, although we find some
hybrid equilibria for low connectivities,  particularly in ER graphs with
$m = 5$ and in BA scale free graphs.
Levels of frustration go up since every agent now coordinates
better with the neighbors, and this takes the final configuration to more
frequent specialized equilibria.
\begin{figure}[h]
{\centering
(a)\subfloat{\includegraphics[width=3in]{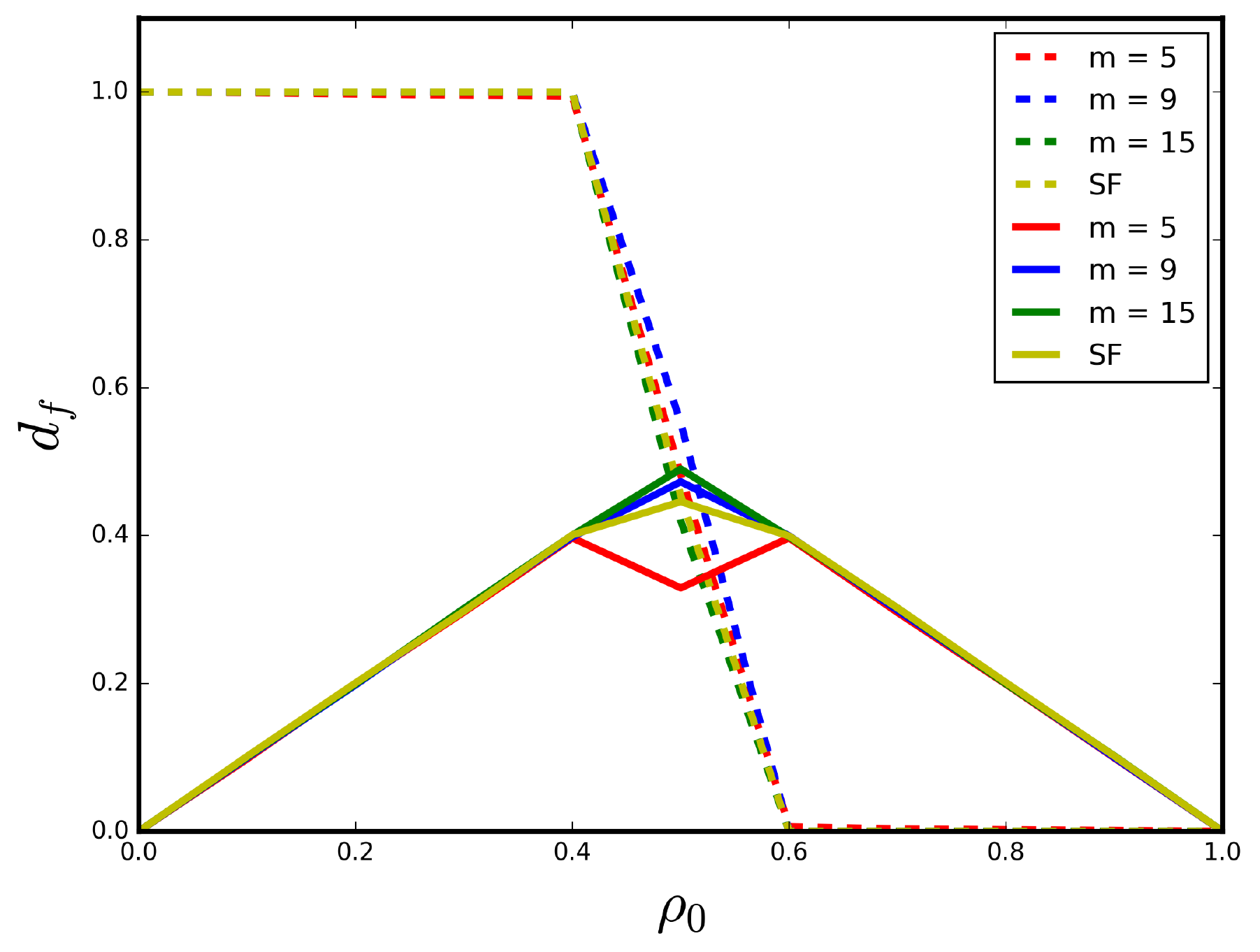}}
(b)\subfloat{\includegraphics[width=3in]{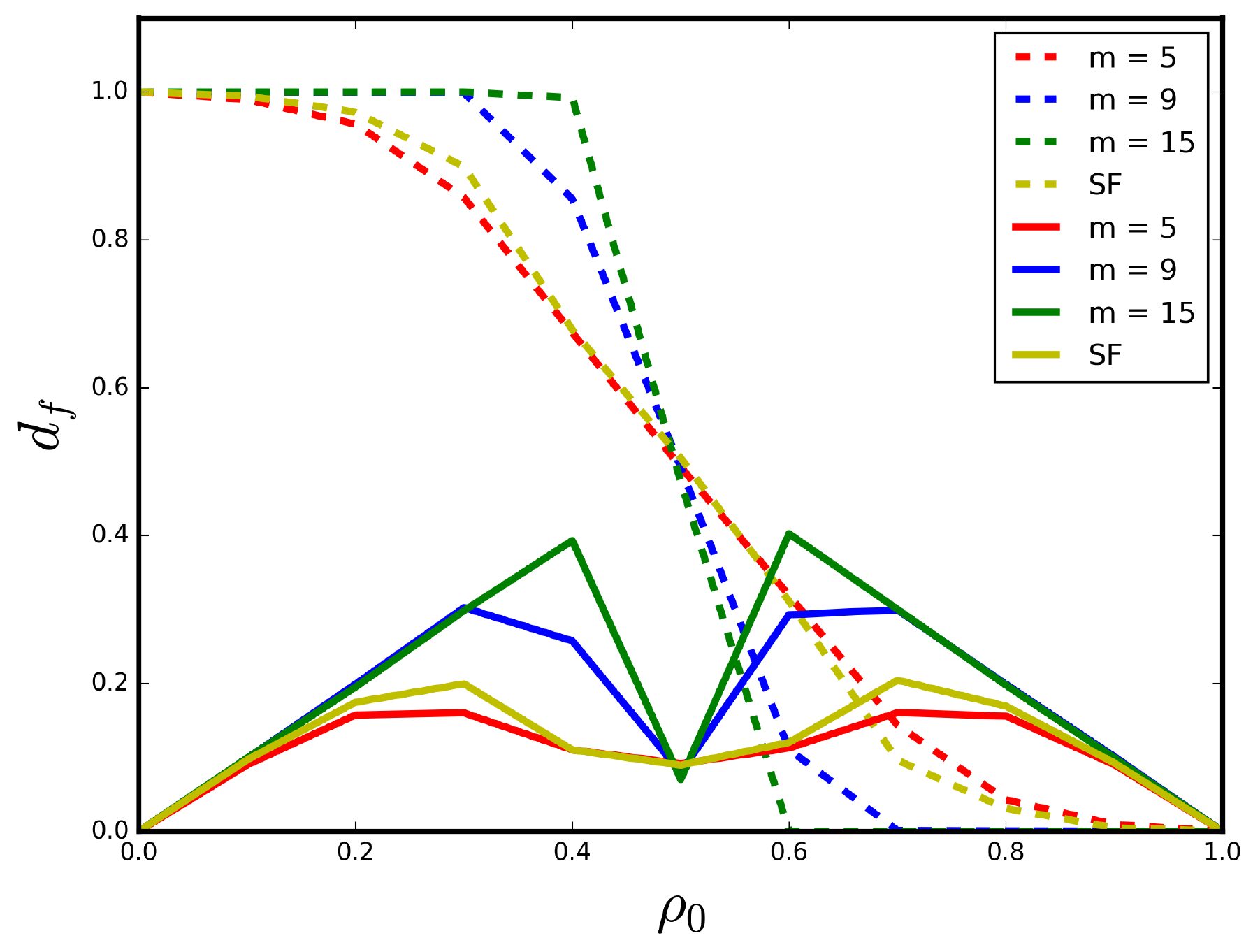}}
}
\caption{Final average density of frustrated agents $d_f$ over
50 realizations against 0-preference density $\rho_0$ (solid lines). Shown also is the corresponding final average density $d_1$ (dashed lines). (a) Coordination Game with reward ratio $\alpha/\beta=1$,  (b)
Coordination Game with reward ratio $\alpha/\beta=2$.}
\end{figure}
Figure 5 confirms this interpretation: Indeed, for $\alpha/\beta=2$
we observe a small difference from the case with 100 nodes, with 
a phase transition from a specialized equilibrium to the other
that is much sharper for high connectivities. Correspondingly, 
frustration curves in figure 5 show higher curves than before since now coordination
is more frequent and agents prefer to change action when $\alpha/\beta=1$.
On the contrary, when $\alpha/\beta=2$ preference matters more when preference
distribution is close to equal compositions, and in fact when
the distribution is 50-50 the system reaches a
more satisfactory equilibrium even if it is not hybrid satisfactory. Therefore, what we observe is that for larger systems coordination is found in a wider range of fractions of 0-preference agents, but that even for 1000 agents there is still quite a sizable range for which hybrid equilibria are possible. 

\subsubsection{Proportional Imitation}

We now turn to the study of the model under proportional imitation dynamics. In this case, 
we simulated ER random graphs for different values of
connectivities and a BA scale free graph with only
10 iterations to save computing time, because reaching the equilibrium
takes sometimes much longer times than in the deterministic case of best response discussed in the previous subsubsection.  
\begin{figure}[h]
{\centering
(a)\subfloat{\includegraphics[width=3in]{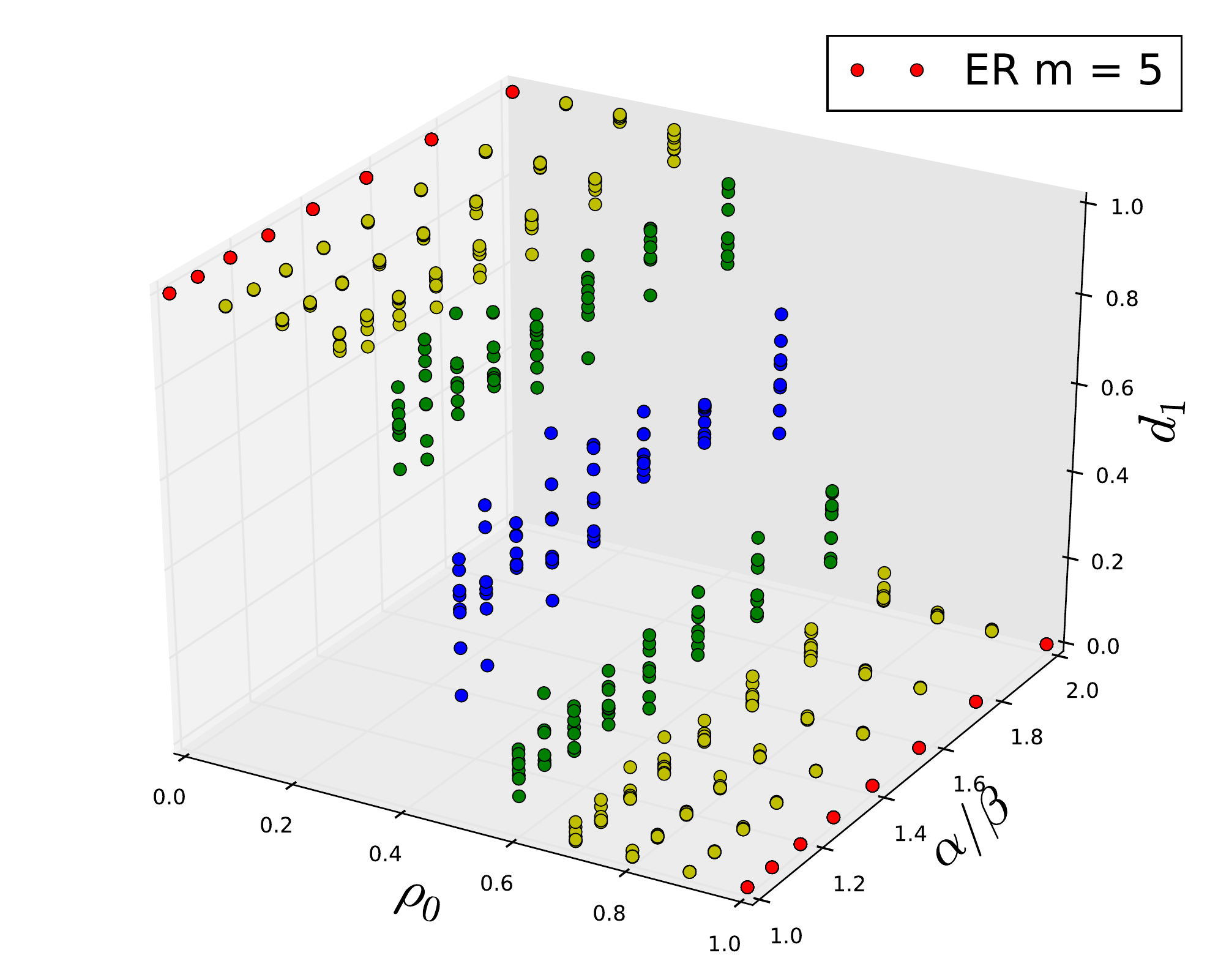}}
(b)\subfloat{\includegraphics[width=3in]{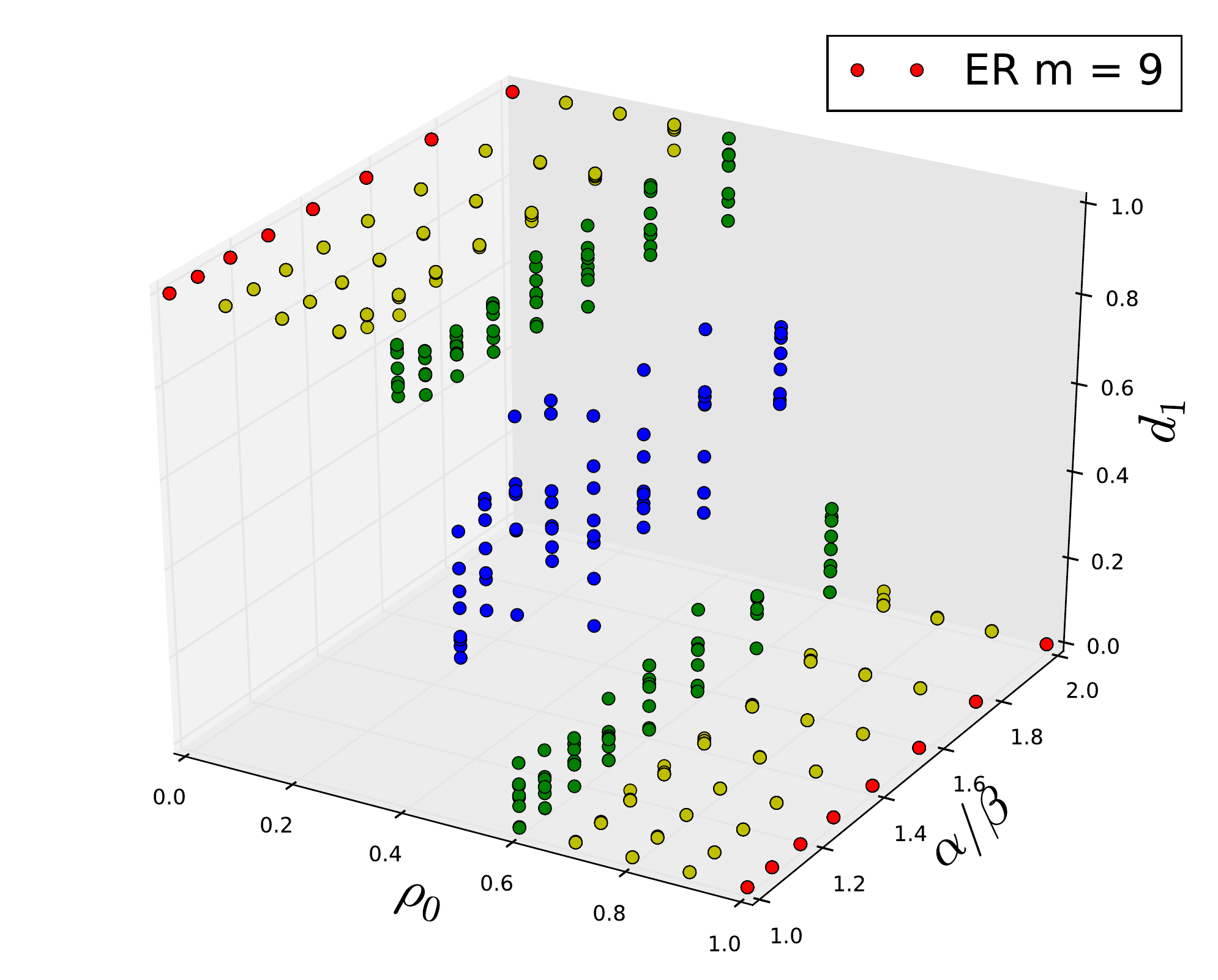}}\\
(c)\subfloat{\includegraphics[width=3in]{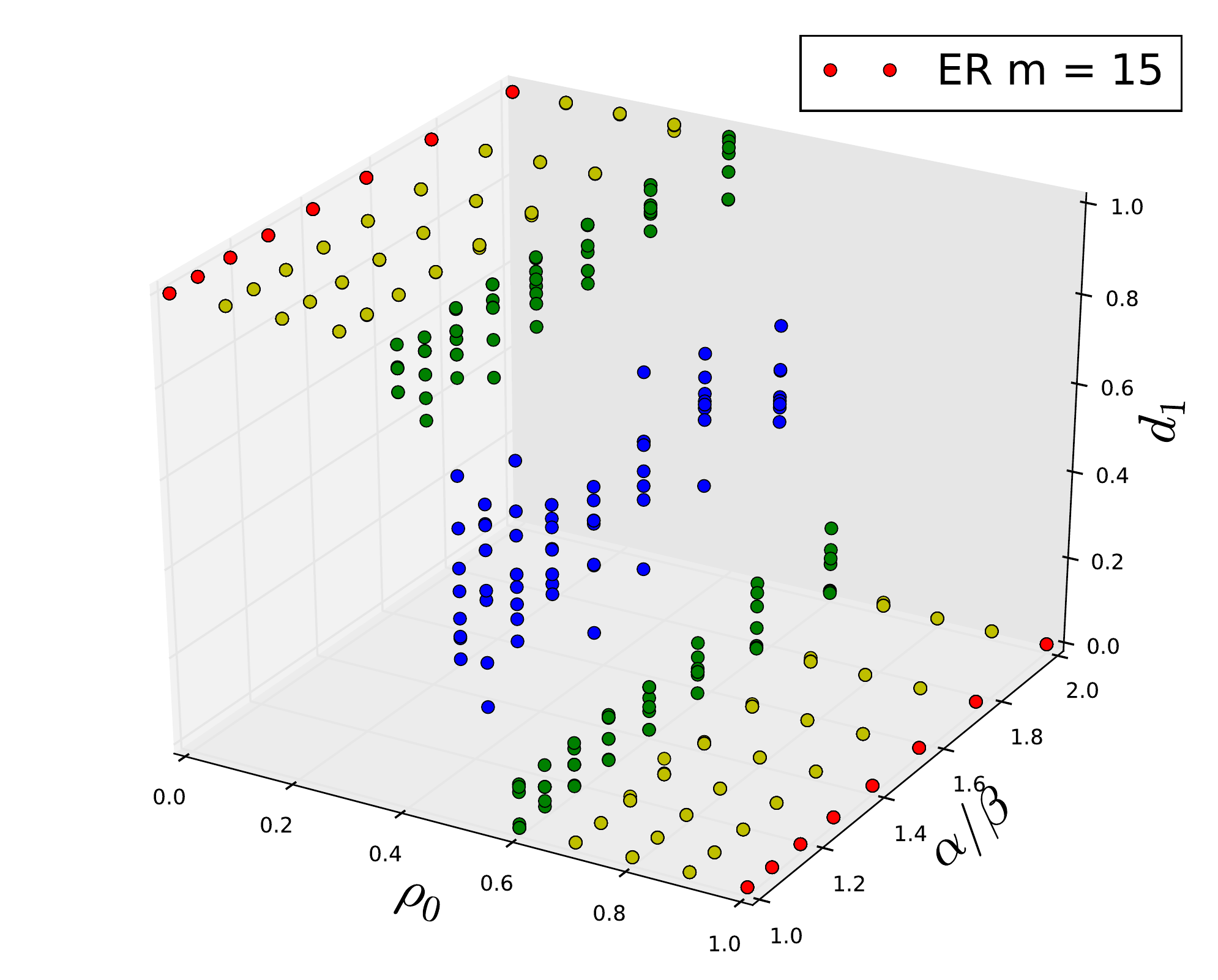}}
(d)\subfloat{\includegraphics[width=3in]{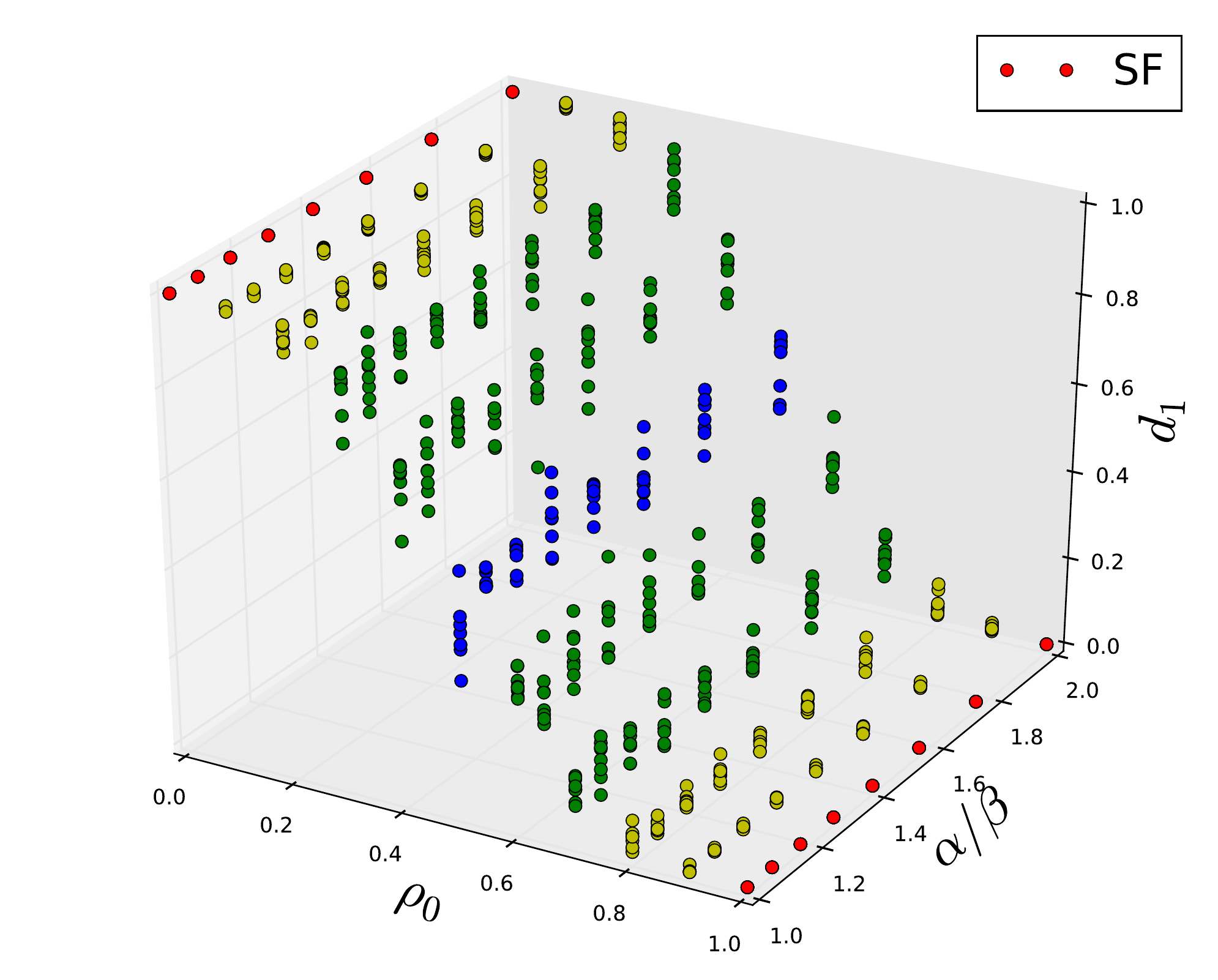}}
}
\caption{Final average density of agents who play action 1 $d_1$ against fraction of 0-preference players $\rho_0$ and reward ratio $\alpha/\beta$ in equilibrium, for the CG on different ER networks (connectivity as indicated in the plot) and a BA network. Colors as in figure 2.}
\end{figure}
It is interesting to keep in mind that in simulations of the homogeneous model with proportional imitation no hybrid
equilibria were found \cite{us:2014}, the only equilibria arising being specialized. In our study, for
the heterogeneous model hybrid equilibria do appear,
expecially in the scale free graph as shown in figure 6. While in the ER random graphs
hybrid equilibria appear only in the neighbourhood of a 50-50 distribution
of preferences, in the scale free graphs almost the whole set of parameters
leads to the emergence of hybrid equilibria.

Frustration curves (figure 7) do not show large differences between
ER and BA graphs, the main reason being that when the reward
ratio is low and the distribution is equal, the selection
dynamic goes totally random. This is so because agents choose a random
neighbor, independently of her preference, and subsequently they choose their action if the
payoff is better than their own one. When $\alpha/\beta=1$ this implies that half of the
0-individuals and the half of the 1-individuals will eventually change
their action, which results in a 50\% of frustration in the final
state. This is so independently of the type of network because for this dynamics agents update their action without taking into account their whole neighborhood as they only look at a randomly chose neighbor. 
\begin{figure}[h]
{\centering
(a)\subfloat{\includegraphics[width=3in]{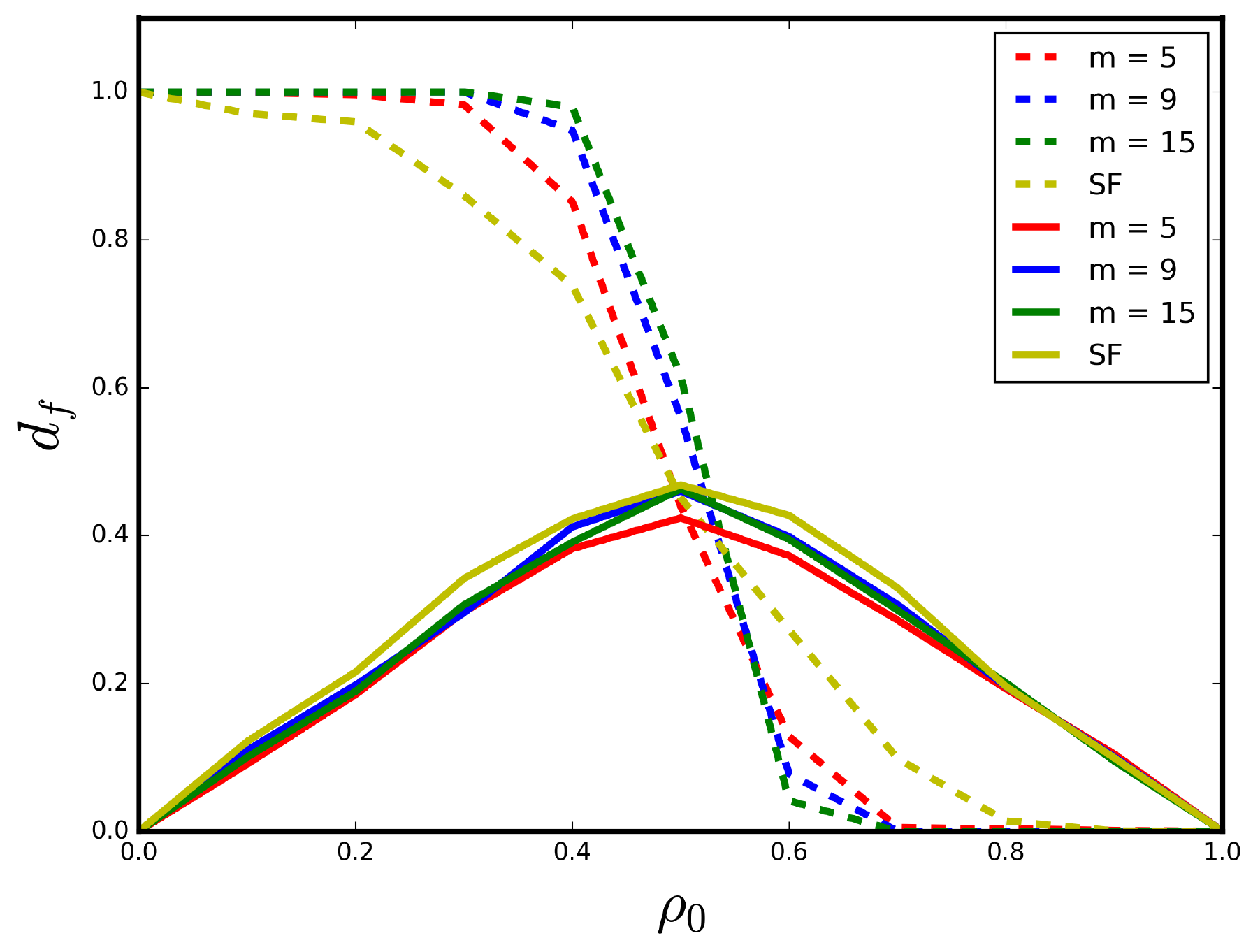}}
(b)\subfloat{\includegraphics[width=3in]{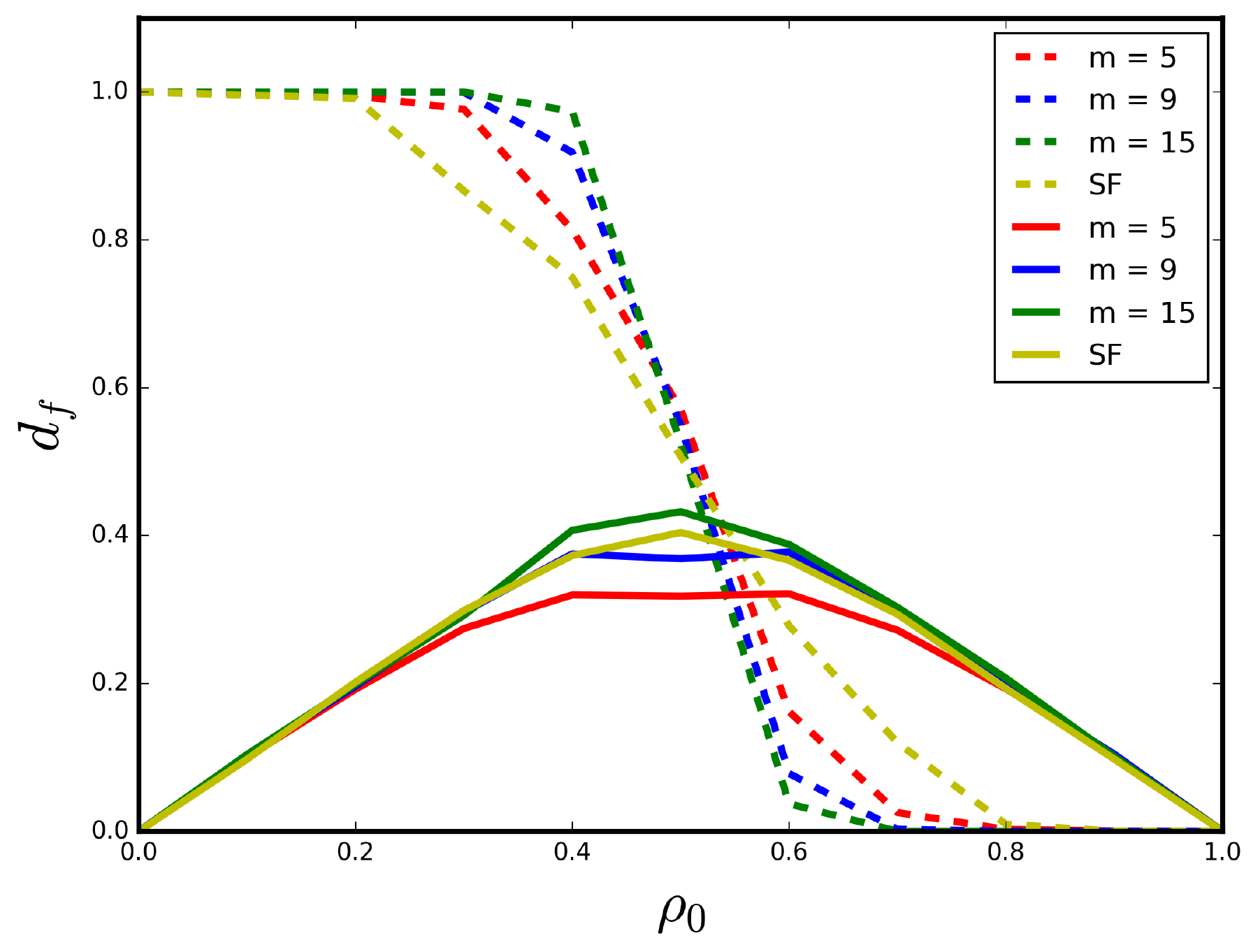}}
}
\caption{Final average density of frustrated agents $d_f$ over
10 realizations against 0-preference density $\rho_0$ (solid lines). Shown also is the corresponding final average density $d_1$ (dashed lines).  Coordination Game with
reward ratio $\alpha/\beta=1$,   Coordination Game with
reward ratio $\alpha/\beta=2$.}
\end{figure}
On the other hand, the reward ratio does not affect the sharpness of the crossover from one specialized equilibrium to the other one
as the connectivity does: Indeed, less connected ER and BA scale free
graphs in figure 7 show more smooth crossovers for both values of the reward ratio, while the frustration curves are also very similar for $\alpha/\beta=1$ and $\alpha/\beta=2$. 

\subsection{Anticoordination Game}

\subsubsection{Best Response}

In this subsection, we will be dealing with AG, i.e., strategic interactions in which the best thing to do is the opposite of one's partners.  However, this is not easy in so far as in our model players intrinsically prefer a specific action over the other, which may coincide with that of their partners. In our simulations for the AG
we do not observe very relevant differences for
different connectivities, but in figure 8 we do observe differences between ER
and BA scale free graphs: For the former, the dependence on the reward ratio is more smooth, but for the BA network (and, to some extent, for the ER network with $m=5$) it appears that there is a type of behavior when 
$\alpha/\beta\lesssim1.5$, and a different one for larger values.  Small reward ratios lead to behavior that is mostly independent of the preference composition of the population, whereas larger reward ratios give rise to final states in which there is a linear relation between the density of 0 actions and the density of 0 agents. In other words, for large $\alpha/\beta$ less agents will feel inclined to change their preferred action to anti-coordinate with their neighbors. Figure 9, that shows the frustration dependence on the composition for the two extreme cases of the reward ration, indicates clearly that this is the case. Another interesting feature that this plot shows is that, opposite to the case of CG, the minimum frustration occurs for intermediate compositions, being more clear for large $\alpha/\beta$. 
Specialized equilibria do not exist in this case for any population composition, and neither do satisfactory
equilibria. Interestingly, for low reward ratios anti-coordination is almost perfect, in the sense that half the agents choose one action and the  other half choose the other, but their choices do not correlate with their preferences, which in turn makes half the population frustrated. 

Comparing this case with the same one in the homogeneous model is
not easy since we had no reward parameters in that case. 
In the homogeneous model connectivity fostered defection, whereas in the heterogeneous
model connectivity has no role as we have just discussed. Similarities can be seen
when we take into consideration scale free graphs, because in this
structures anticoordination works generally better and frustration
is reduced in heterogeneous distributions. In fact, in the homogeneous
model, as in the heterogeneous one we see that final configurations
are better anticoordinated than those in the random graphs.
\begin{figure}[h]
{\centering
(a)\subfloat{\includegraphics[width=3in]{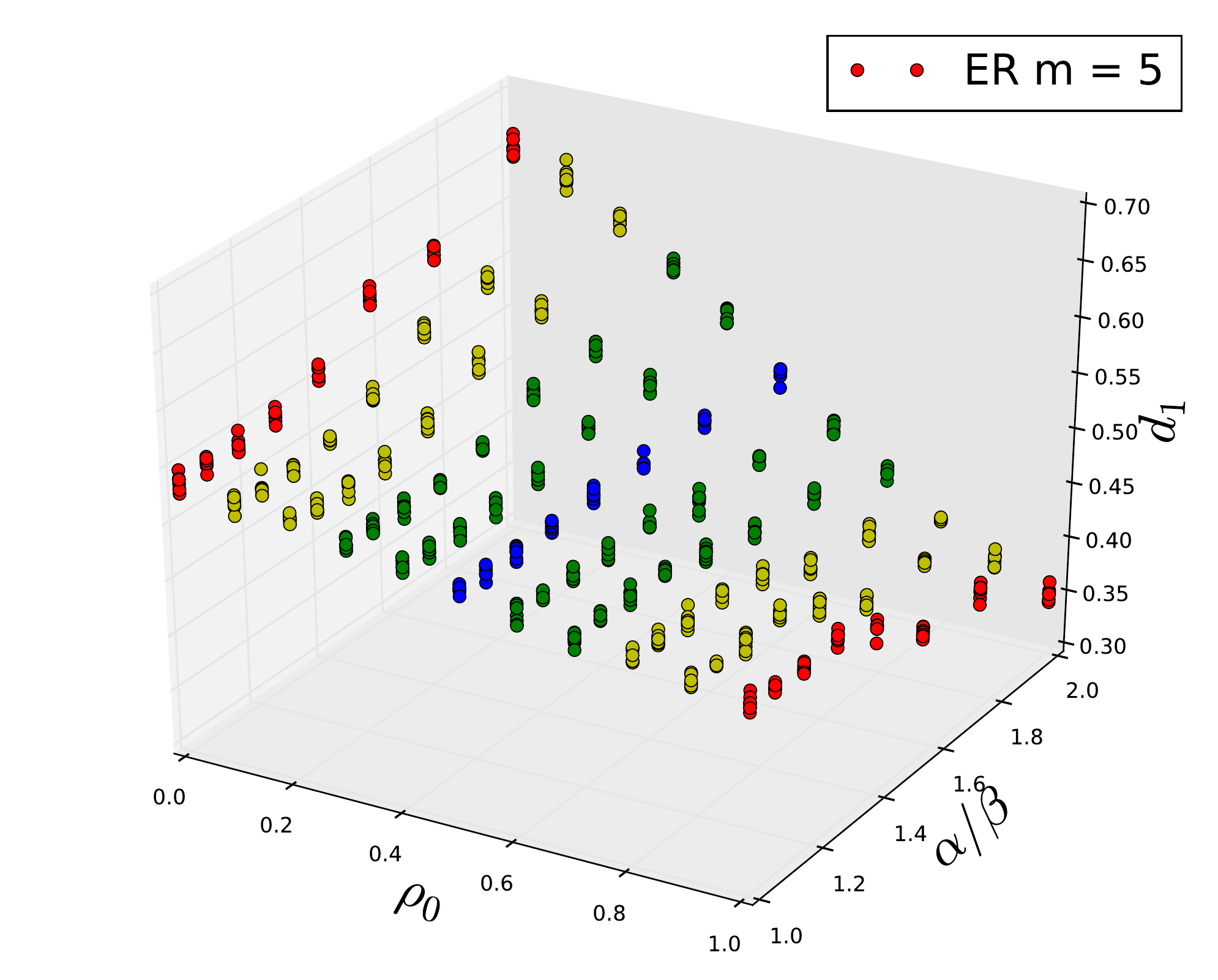}}
(b)\subfloat{\includegraphics[width=3in]{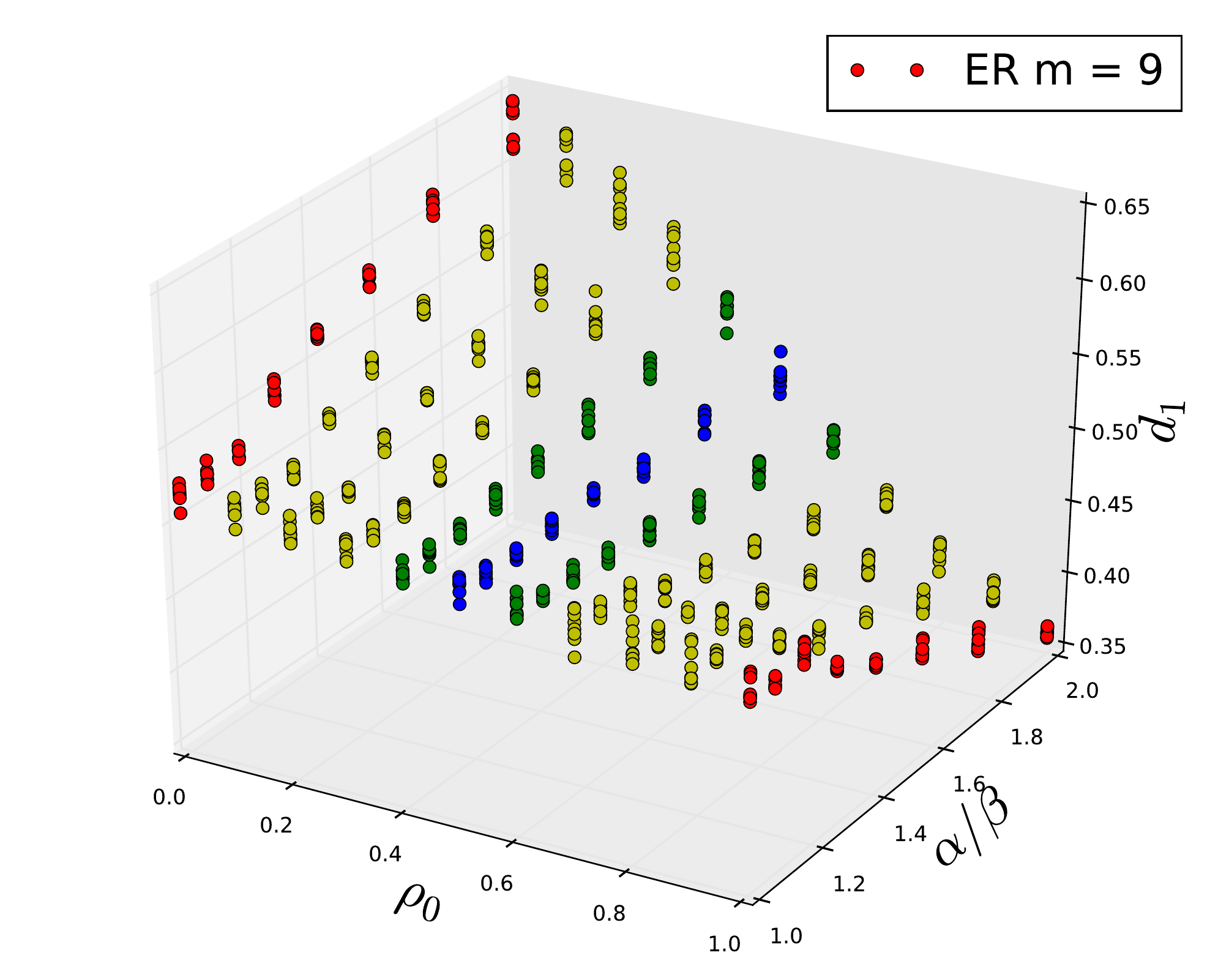}}\\
(c)\subfloat{\includegraphics[width=3in]{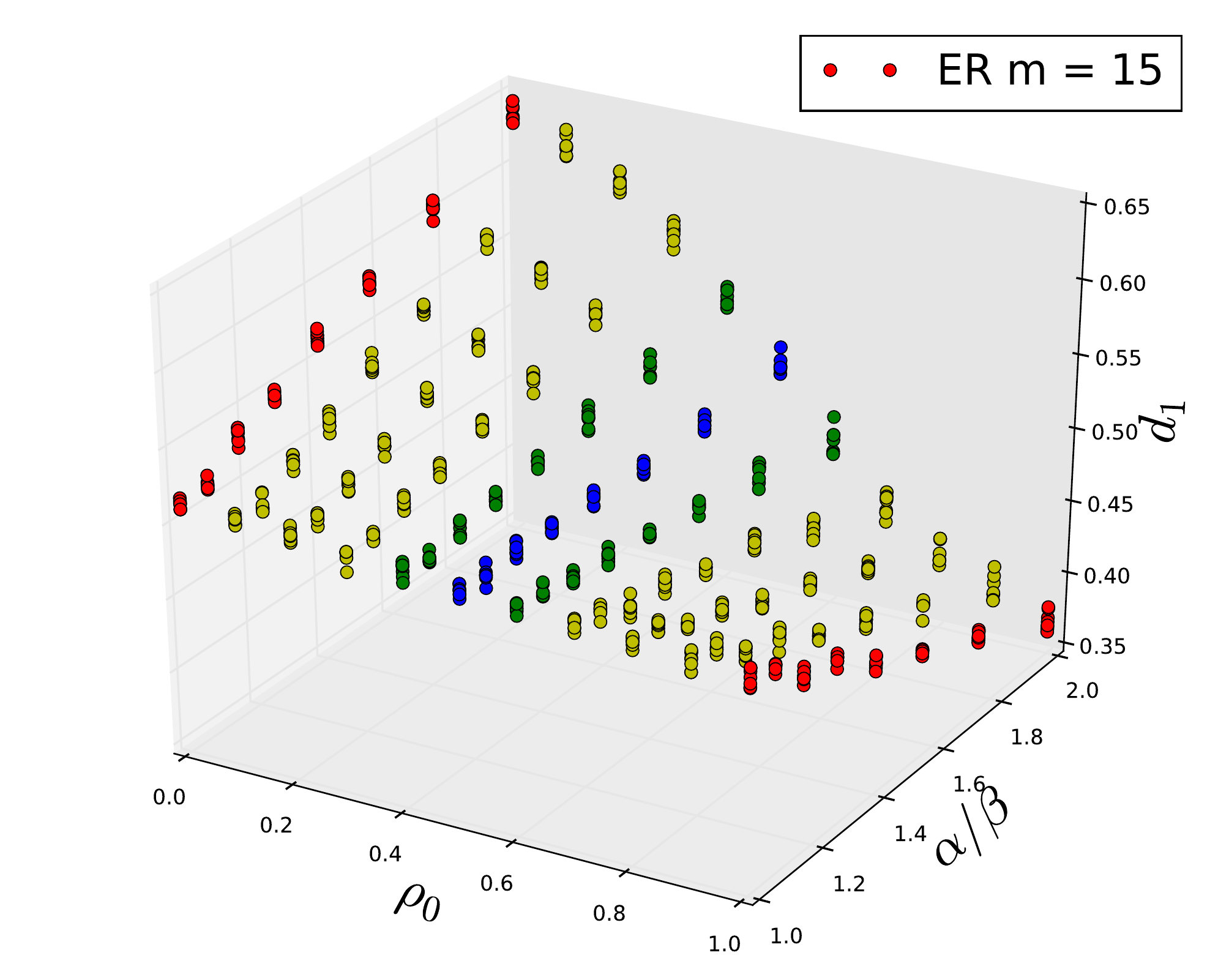}}
(d)\subfloat{\includegraphics[width=3in]{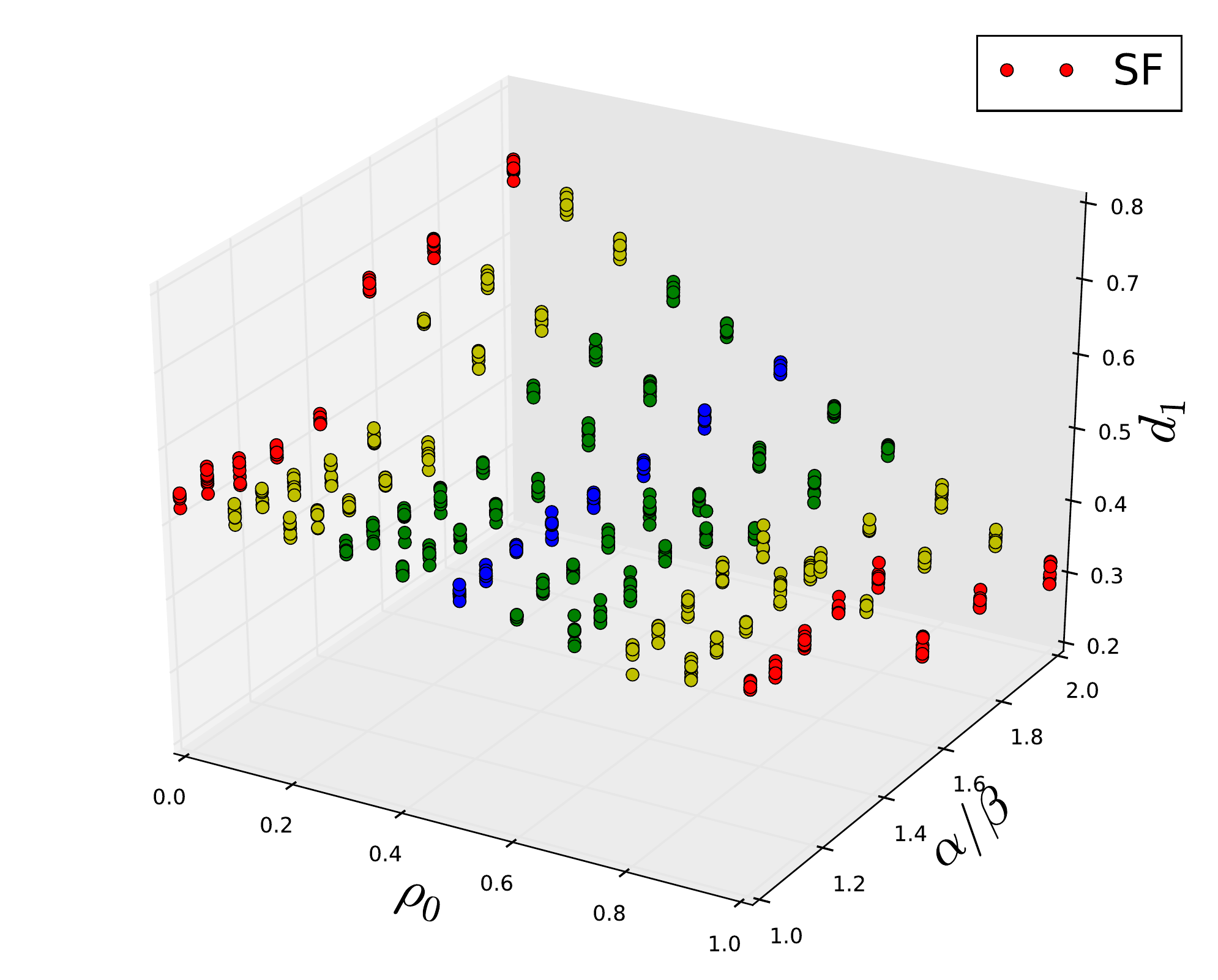}}
}
\caption{Final average density of agents who play action 1 $d_1$ against fraction of 0-preference players $\rho_0$ and reward ratio $\alpha/\beta$ in equilibrium, for the AG on different ER networks (connectivity as indicated in the plot) and a BA network. Colors as in figure 2.}
\end{figure}
\begin{figure}[h]
\centering
(a)\subfloat{\includegraphics[width=3in]{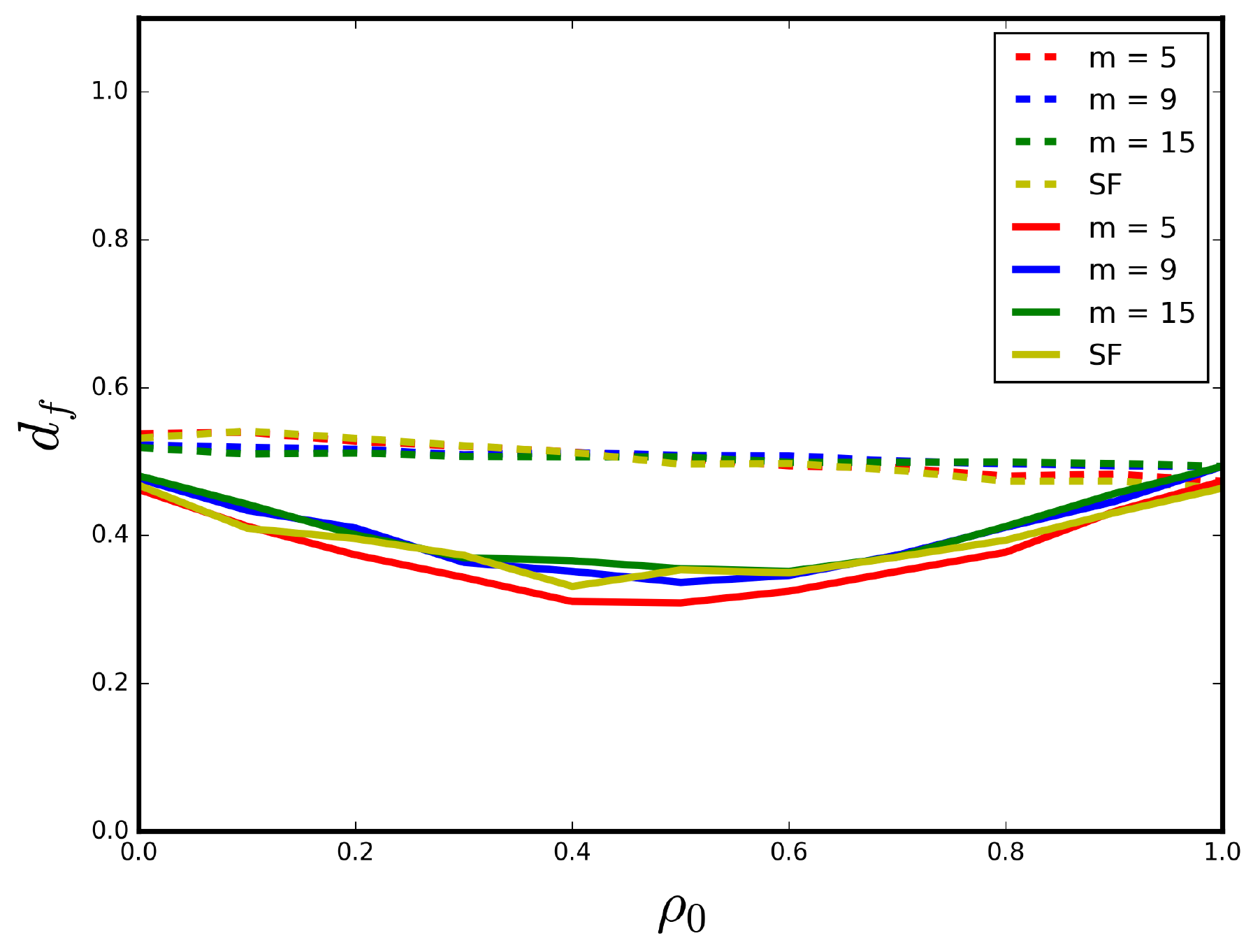}}
(b)\subfloat{\includegraphics[width=3in]{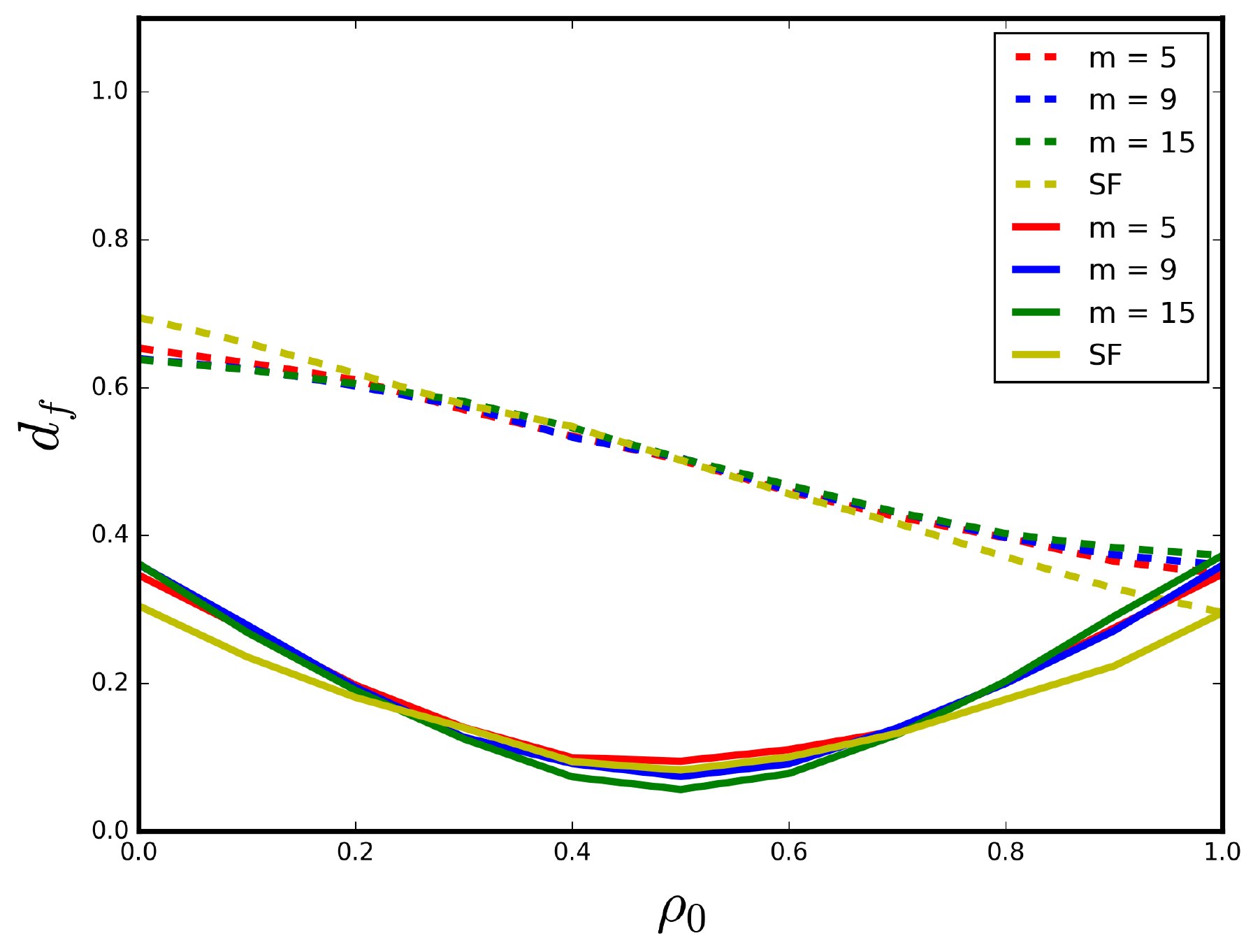}}\\
\caption{Final average density of frustrated agents $d_f$ over
50 realizations against 0-preference density $\rho_0$ (solid lines). Shown also is the corresponding final average density $d_1$ (dashed lines).  a) Anti-coordination Game with
reward ratio $\alpha/\beta=1$, b) Anti-coordination Game with
reward ratio $\alpha/\beta=2$.}
\end{figure}

\subsubsection{Proportional Imitation}

The dynamics of AG under proportional imitation is, generally speaking, similar to that under best response, but there are some specific features worth discussing. 
First of all, with a homogeneous distribution the system cannot change
his state by definition, since no agent can imitate an action that
nobody is playing. This is represented by the extreme cases (red dots) in figure 10. Outside these special values, 
a large degree of anticoordination is achieved in most cases. In the case of heterogeneos
preference distributions with reward ratio $\alpha/\beta\rightarrow1$,
anticoordination is reached almost always, except for the
scale free networks, where the transition to specialized equilibria
is smoother than in the random graphs (see also figure 11). As before, 
anticoordination works worse when $\alpha/\beta\rightarrow2$, because
obviously the agents are more motivated to keep on playing their preferred
option.
Frustration final values show that for homogeneous distributions there
is no frustration in the final states, as of course they did not anticoordinate
at all so they kept on playing their liked action till the end. We observed a difference between
the random graphs and the scale free graphs: when $\alpha/\beta\rightarrow1$
frustration is reduced when the distribution is close to 50-50,
but not in the scale free graphs where there is a maximum of frustration;
when $\alpha/\beta\rightarrow2$ it appears that scale free architecture makes it
more difficult to anticoordinate and to stay satisfied.
With $\alpha/\beta=1$ the crossover to specialized equilibria
is sharp, although not so much for low connectivity graphs. On the
contrary with a high reward ratio $\alpha/\beta=2$, connectivity does
not affect at all the equilibria, but raising the reward ratio makes
the crossover much smoother than before.
\begin{figure}[htb]
{\centering
(a)\subfloat{\includegraphics[width=3in]{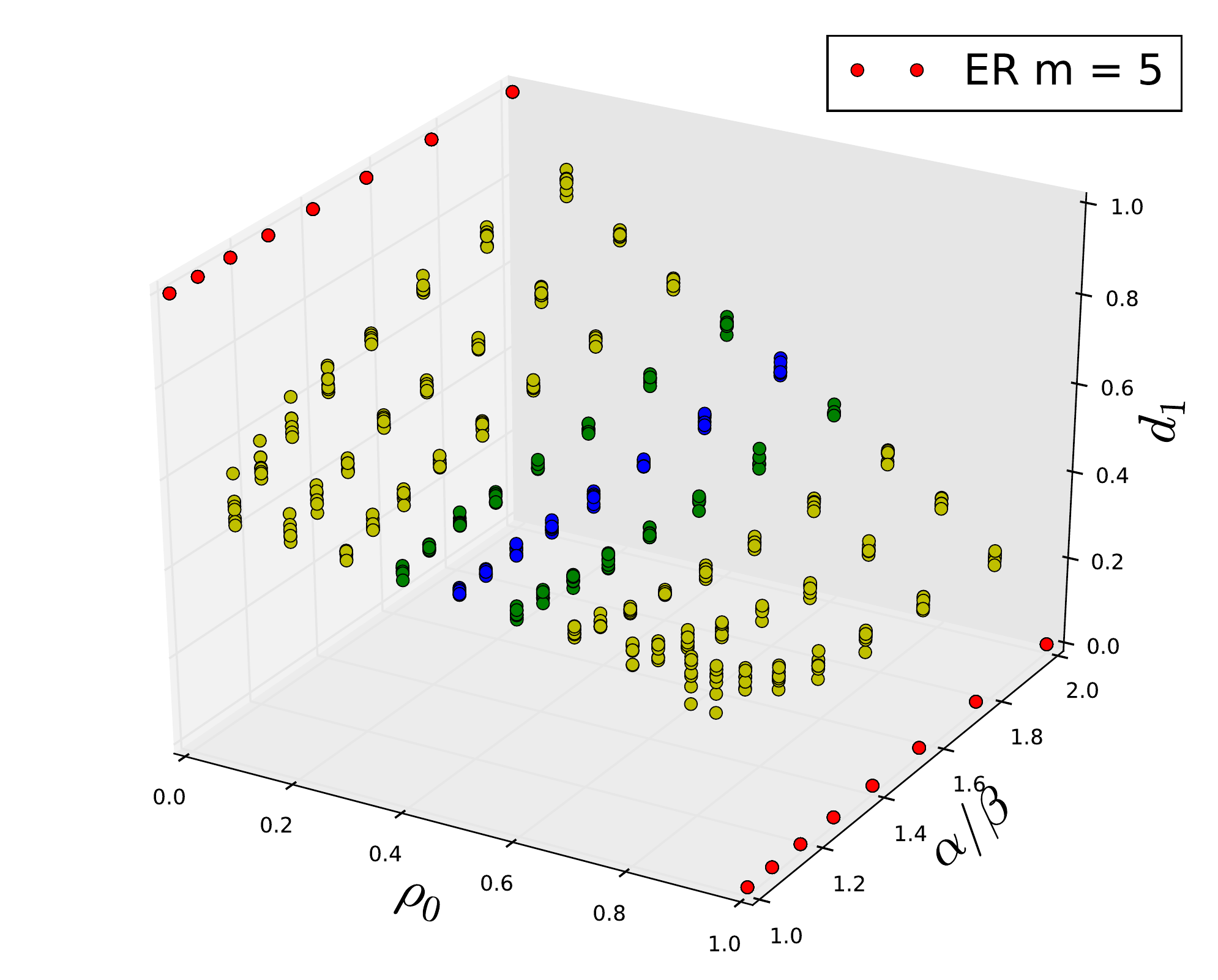}}
(b)\subfloat{\includegraphics[width=3in]{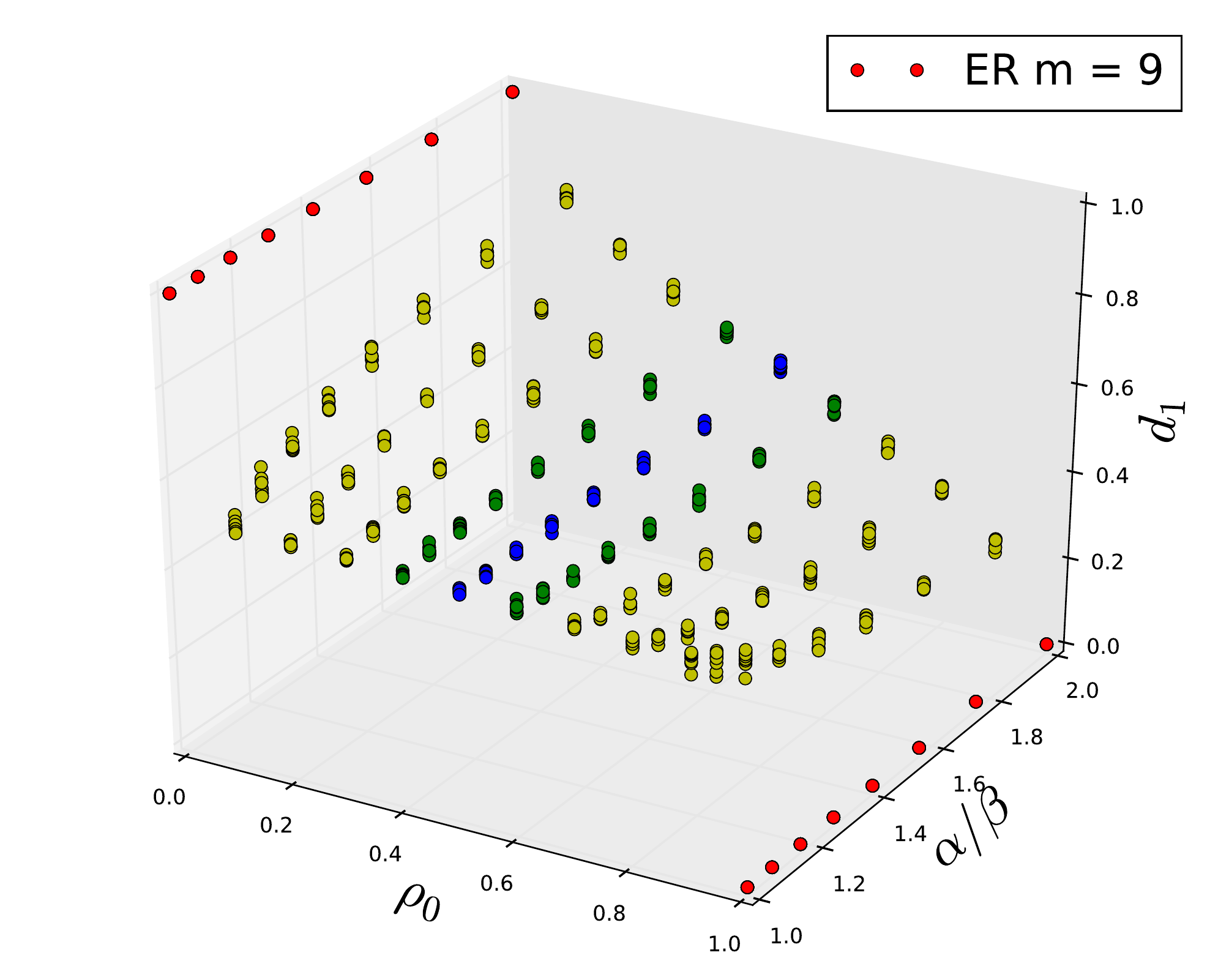}}\\
(c)\subfloat{\includegraphics[width=3in]{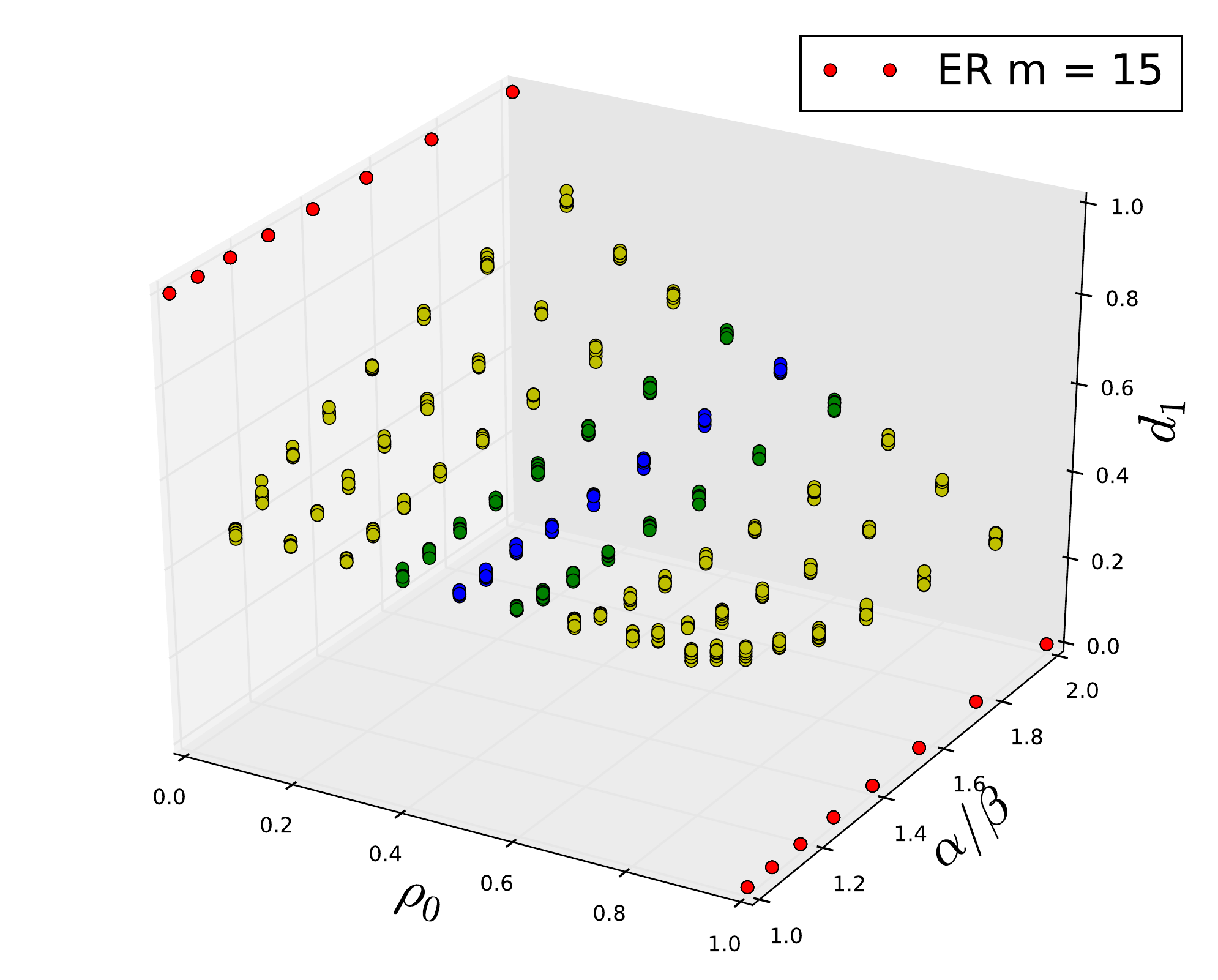}}
(d)\subfloat{\includegraphics[width=3in]{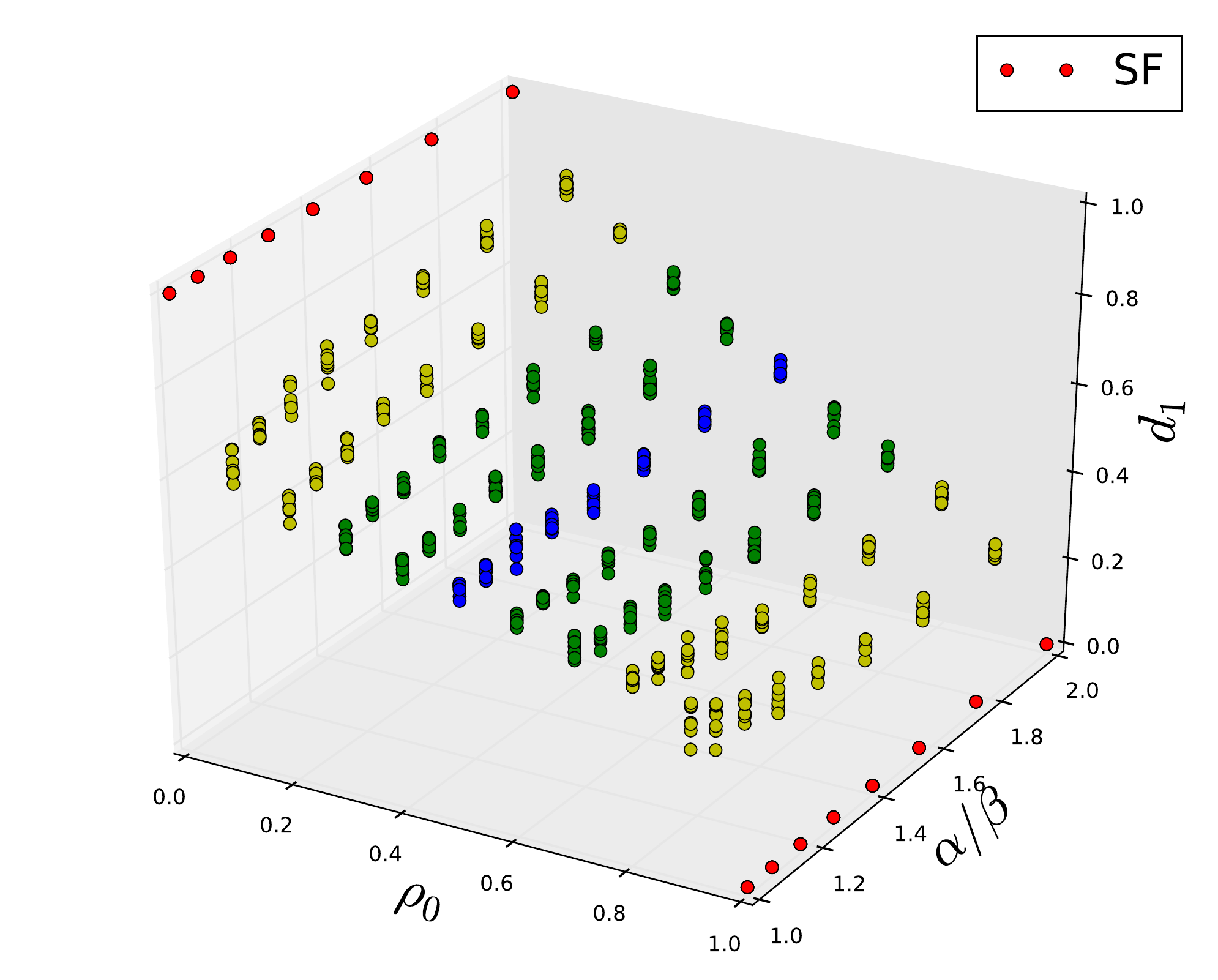}}
}
\caption{Final average density of agents who play action 1 $d_1$ against fraction of 0-preference players $\rho_0$ and reward ratio $\alpha/\beta$ in equilibrium, for the AG on different ER networks (connectivity as indicated in the plot) and a BA network. Colors as in figure 2.}
\end{figure}
\begin{figure}[h]
\centering
(a)\subfloat{\includegraphics[width=3in]{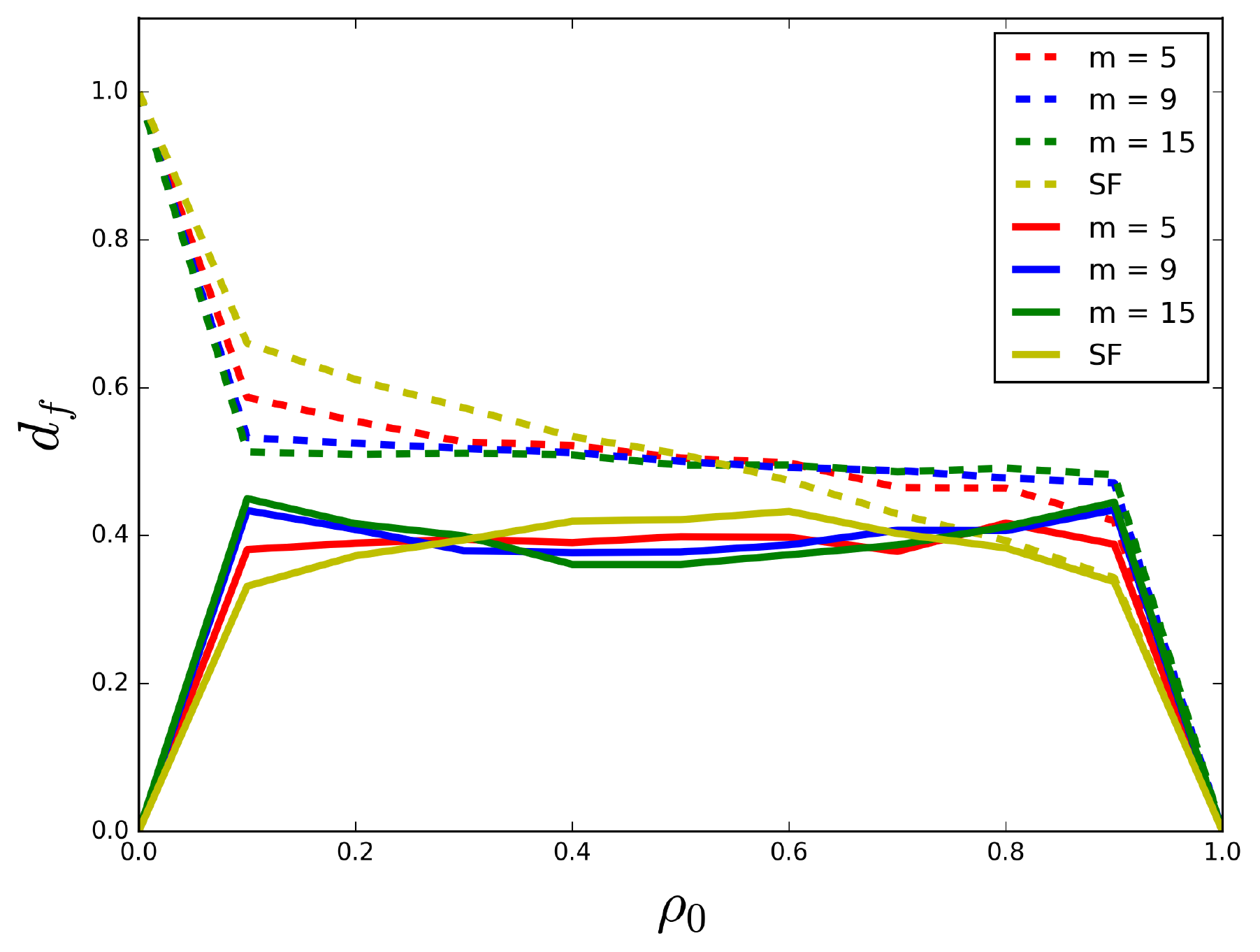}}
(b)\subfloat{\includegraphics[width=3in]{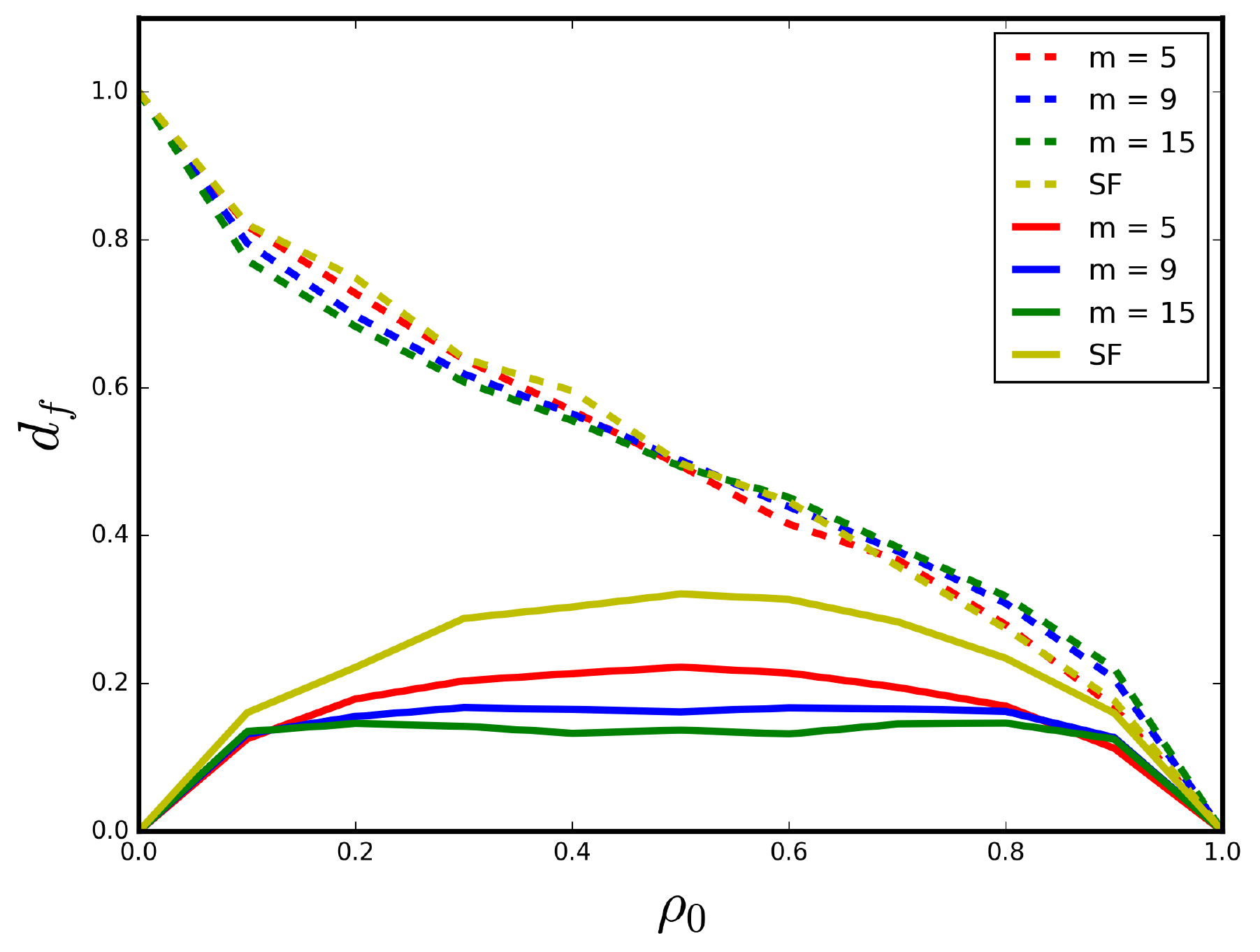}}\\
\caption{Final average density of frustrated agents $d_f$ over
10 realizations against 0-preference density $\rho_0$ (solid lines). Shown also is the corresponding final average density $d_1$ (dashed lines). a) Anti-coordination Game with
reward ratio $\alpha/\beta=1$,  b) Anti-coordination Game with
reward ratio $\alpha/\beta=2$.}
\end{figure}

\section{Incomplete Information}

Thus far, we have been discussing a situation in which all agents have full information about their surroundings, both about types of partners and about their actions. However, in many social contexts it is difficult to have information about others' preferences, and therefore it is worth considering how the results change when we switch to an incomplete information framework. In this case, agents know what they like (i.e., their own preference, of course),
they know how many neighbors they have, but they do not know who
these neighbors are. All they can resort to decide on their action is a distribution of preferences that allows them to estimate the quantity of the two types of neighbors
they will have around them.  This is a quite realistic assumption as very often one has an idea of how opinions or preferences are distributed in the population (e.g., through polls) but is unaware of the specific preferences of the people with whom one is interacting. 
Our aim is to show if the simulation results of our agents based model
fit with the theoretical analysis, which showed how the incomplete information
framework reduces the multiplicity of Nash Equilibrium respect to
those obtained with the complete information framework.

In what follows, we discuss our results under best response dynamics. In the framework of incomplete information, we cannot consider proportional imitation dynamics, since it is not permitted to the agent to know
her neighbour's payoff, the agent knows only her own preference, the
number of neighbours and distribution of preferences present in the
network. Therefore, a payoff comparison is not possible, making the dynamics unapplicable to this case.

\subsection{Coordination Game}

\begin{figure}[h]
{\centering
(a)\subfloat{\includegraphics[width=3in]{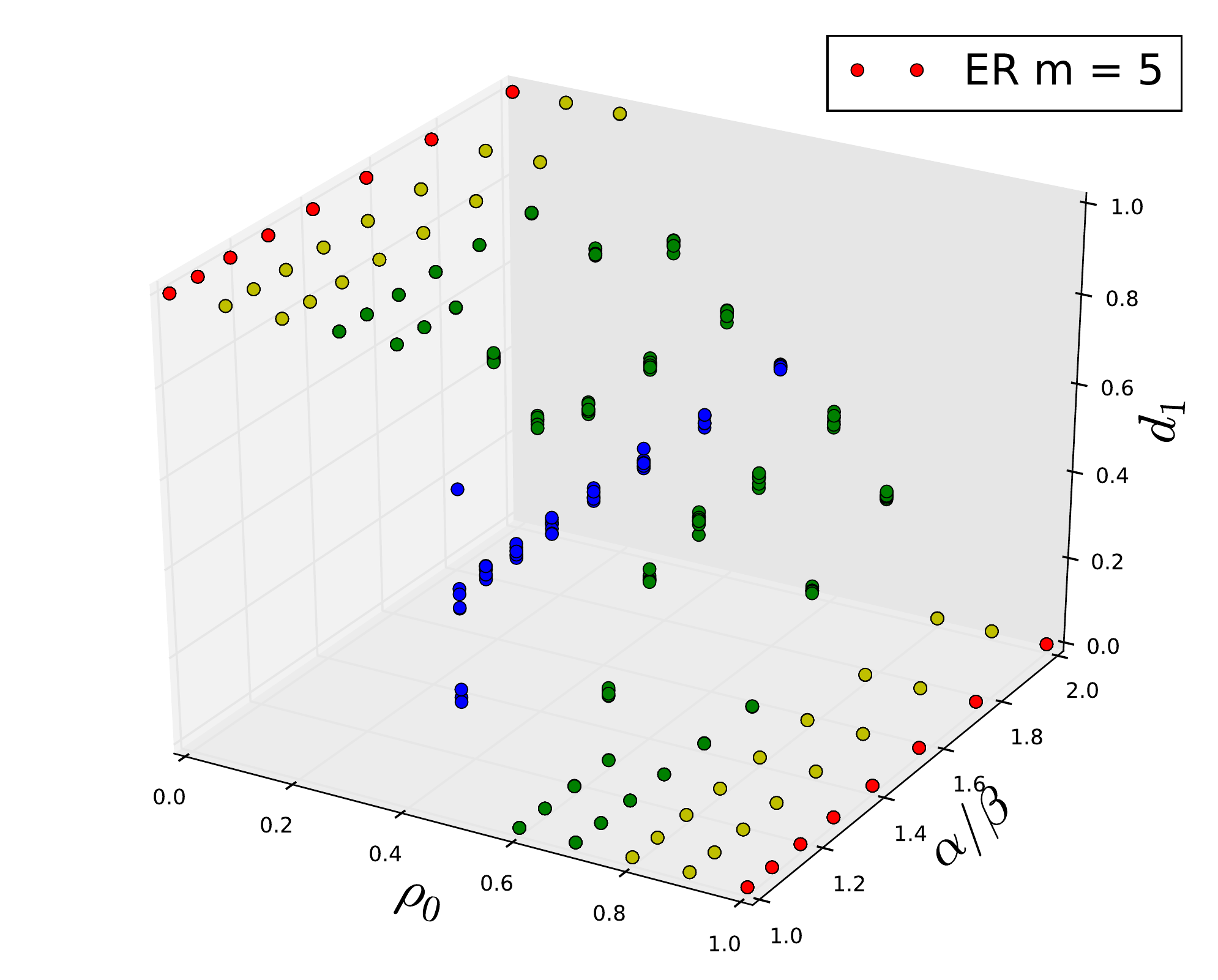}}
(b)\subfloat{\includegraphics[width=3in]{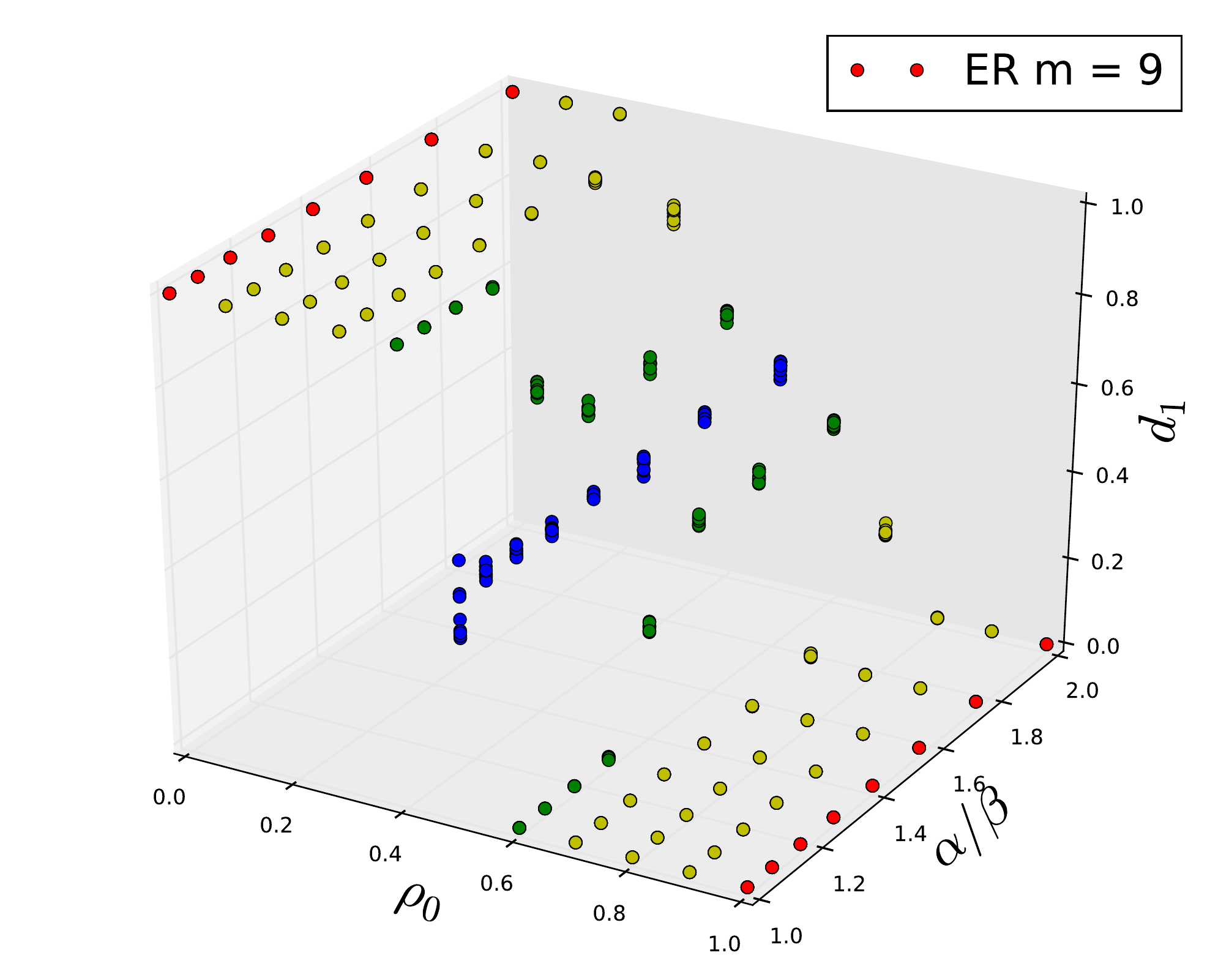}}\\
(c)\subfloat{\includegraphics[width=3in]{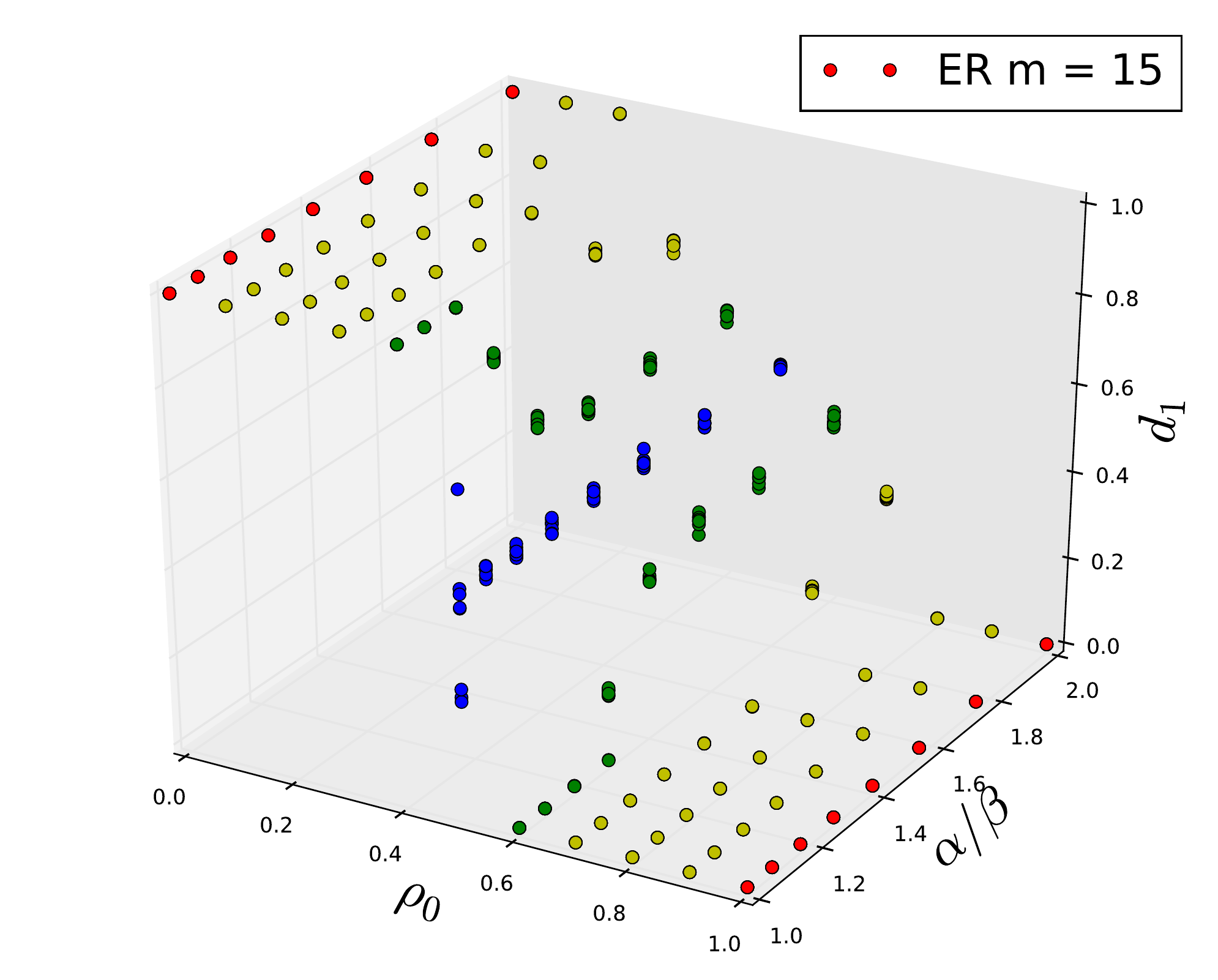}}
(d)\subfloat{\includegraphics[width=3in]{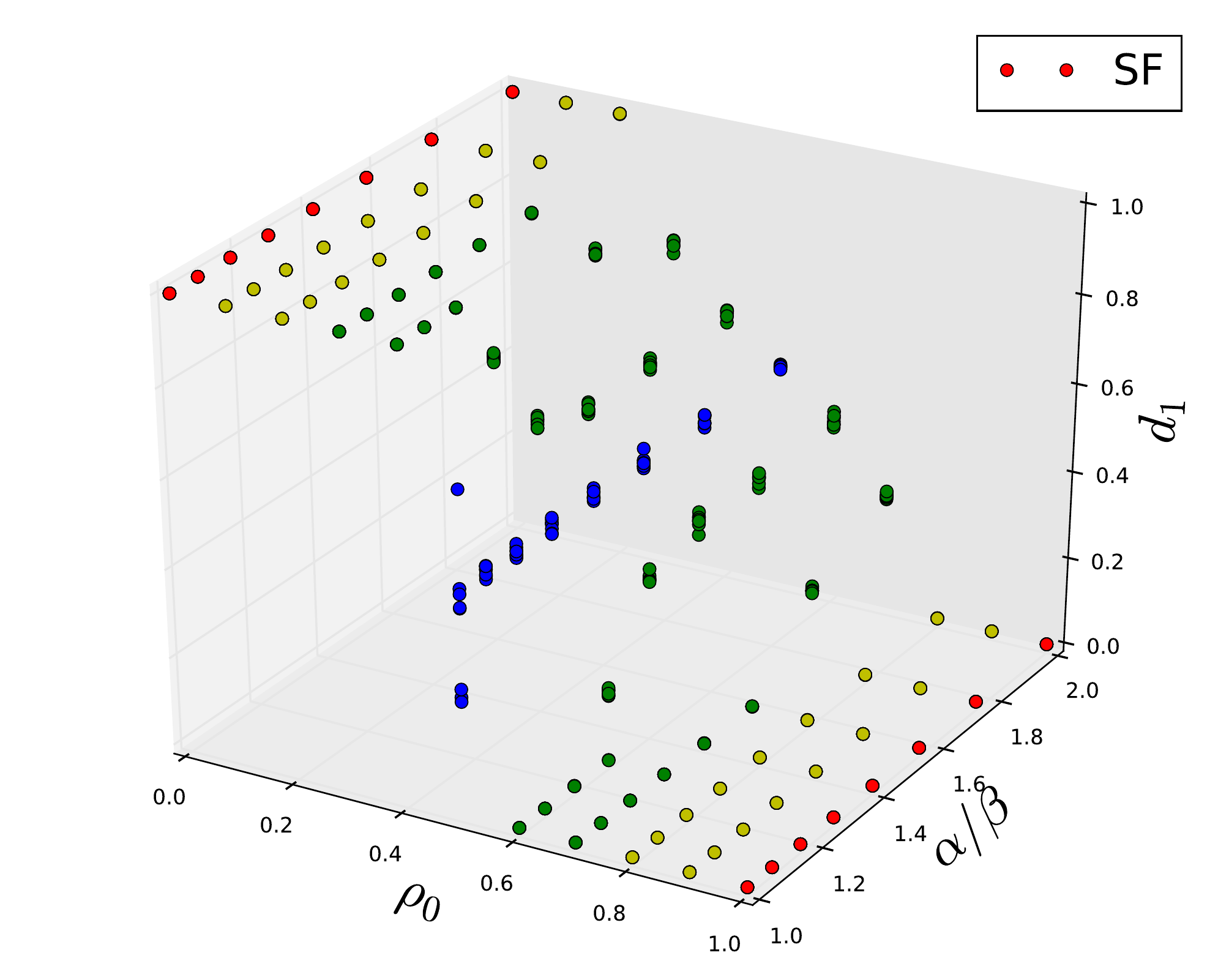}}
}
\caption{Final average density of agents who play action 1 $d_1$ against 0-preference density $\rho_0$ and reward ratio $\alpha/\beta$.}
\end{figure}
Compared with the equilibria we found with the complete information
framework we see a strongly reduced and ordered set of equilibria in figure 12,
confirming what the work of Galeotti et al.\ \cite{key-3} predicted. There are less dots in the figure, indicating that the system ends up in a reduced set of configurations. On the other hand, there are similarities between the two informational setups:
As we discussed above, raising connectivity implies the loss
of many hybrid equilibria, taking the system to more specialized configurations. 
Looking at frustration in figure 13 we see full satisfactory equilibria when we play games with 50-50 distributions. This
agrees with the analytic results obtained \cite{key-2},  where it was found that when the distribution on preferences is
very heterogeneous, $\frac{\alpha}{\alpha+\beta}>\rho>\frac{\beta}{\alpha+\beta}$, with $\rho$ being the fraction of players with preference 1 in the population, 
then satisfactory hybrid configurations appear as a consequence of
symmetric equilibrium. The theoretical predictions are in fact more specific, and can be summarized as follows: There exists a only a pure symmetric equilibrium, and
\begin{itemize}
\item if $\frac{\alpha}{\alpha+\beta}>\rho>\frac{\beta}{\alpha+\beta}$
then every symmetric equilibrium is satisfactory for any connectivity,
\item if $\rho>\frac{\alpha}{\alpha+\beta}$ then the action of a given player 
may only go from 0 to 1 as the degree increases, and all players with preference 1 are satisfied, and
\item if $\rho<\frac{\beta}{\alpha+\beta}$ then the action of a given player may only go from 1 to 0 as the degree increases, and all players with preference 0 are satisfied. 
\end{itemize}

\begin{figure}[h]
\centering
(a)\subfloat{\includegraphics[width=3in]{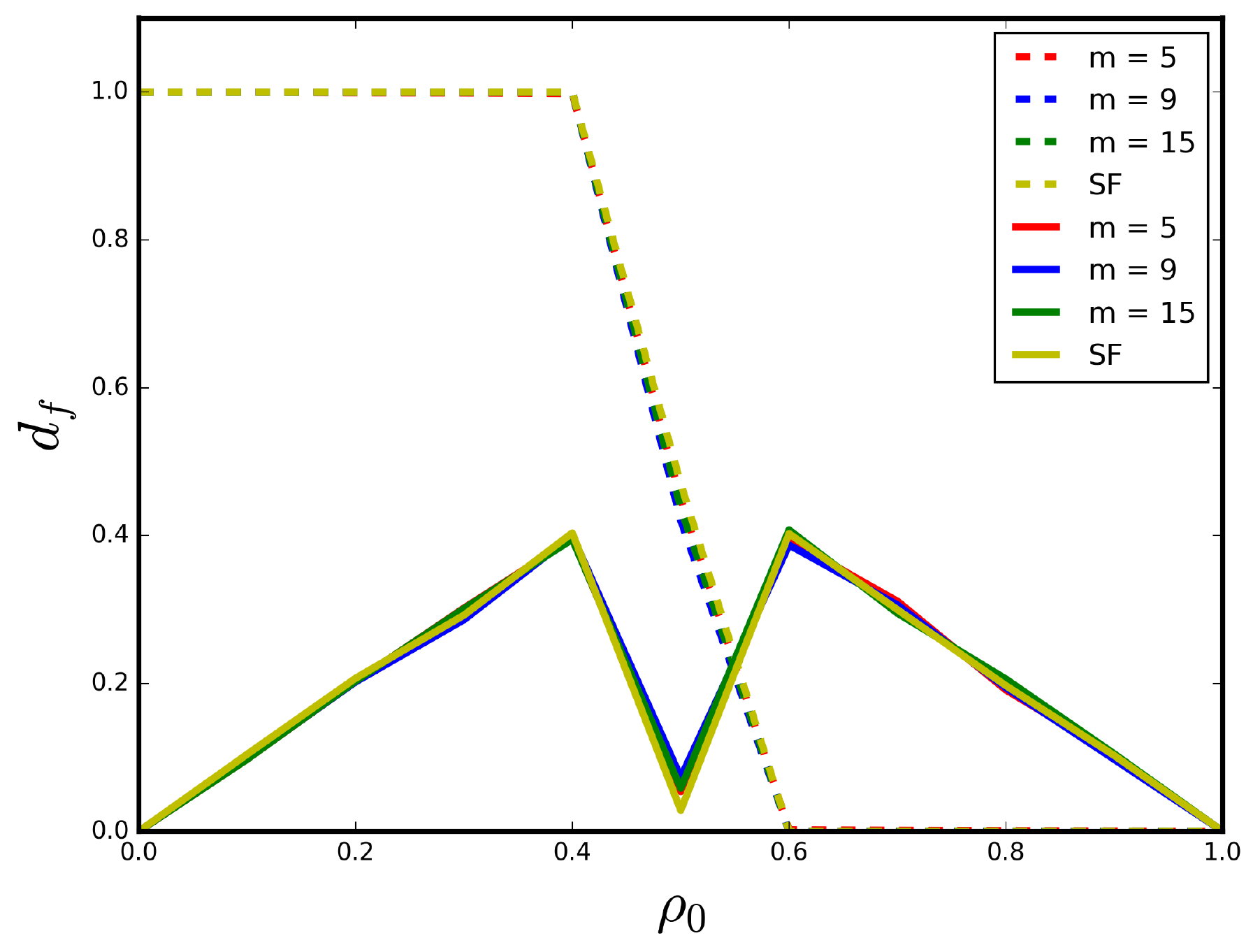}}
(b)\subfloat{\includegraphics[width=3in]{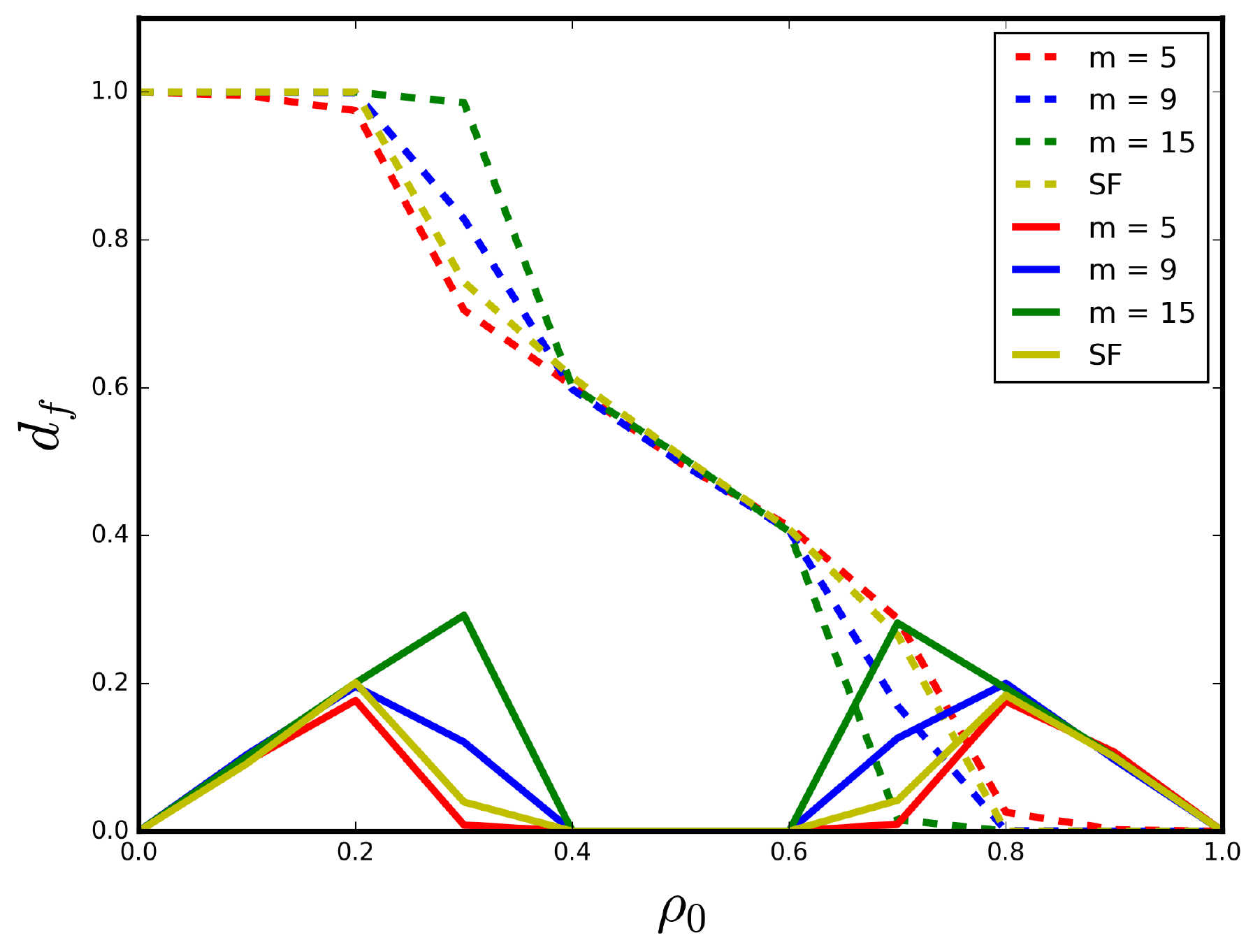}}\\
\caption{Final average density of frustrated agents $d_f$ over
50 realizations against 0-preference density $\rho_0$. \newline
(a) Coordination Game with reward ratio $\alpha/\beta=1$, \newline(b)
Coordination Game with reward ratio $\alpha/\beta=2$.}
\end{figure}
As is also shown in figure 13, similarly to the previous cases, with $\alpha/\beta=1$ connectivity
does not affect at all the sharpness of the crossover, but
for $\alpha/\beta=2$ we notice an interesting linear behavior with respect to the proportion of players of one or the other preference.
With $\alpha/\beta=2$, the plot shows  that in the range of $0.4<\rho<0.6$
full satisfactory hybrid equilibria are reached for the CG in this incomplete information framework.
In figure 14 we show the behaviour of frustration in an alternative presentation to better compare the results directly
with the theory.
\begin{figure}[h]
{\centering
(a)\subfloat{\includegraphics[width=3in]{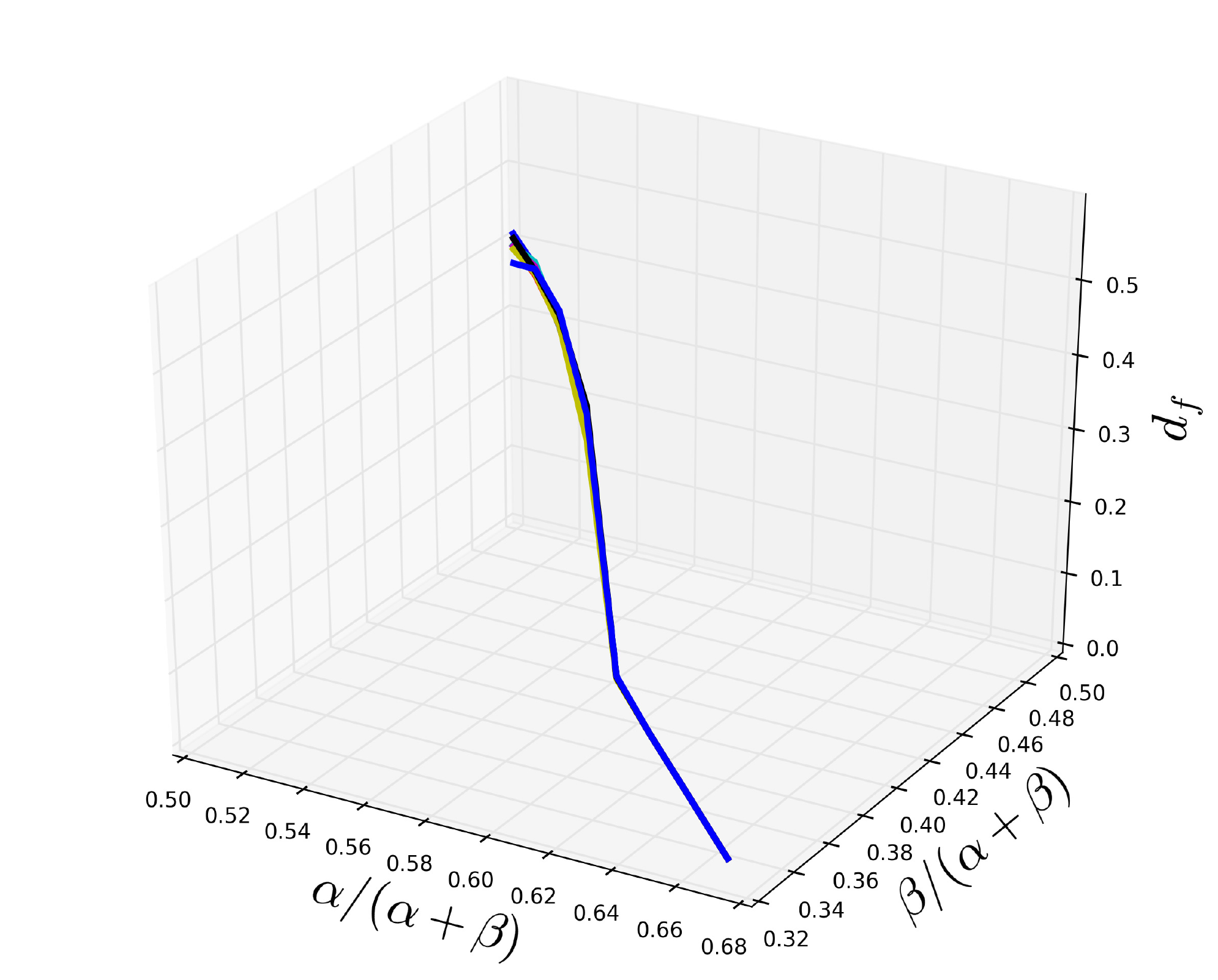}}
(b)\subfloat{\includegraphics[width=3in]{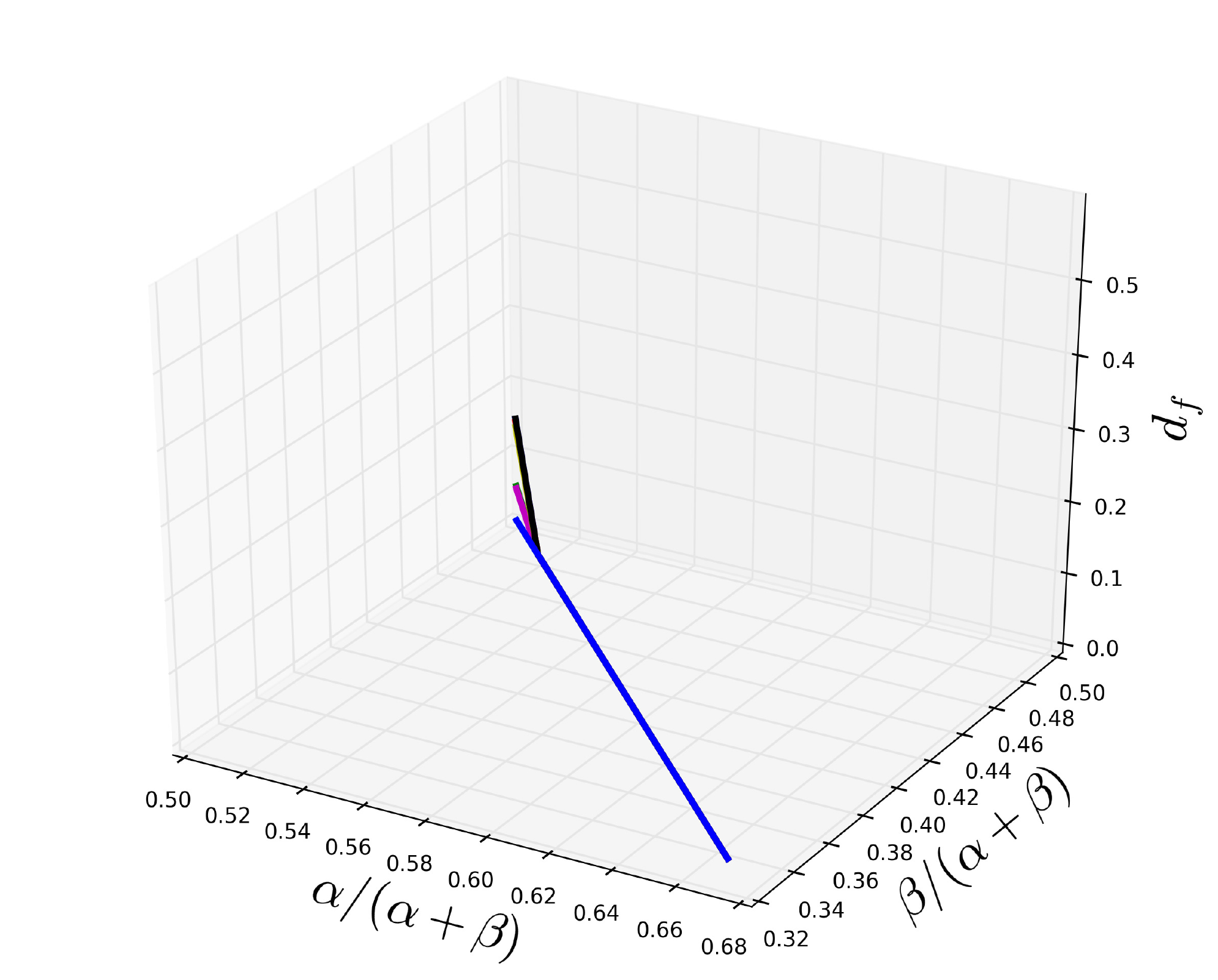}}\\
(c)\subfloat{\includegraphics[width=3in]{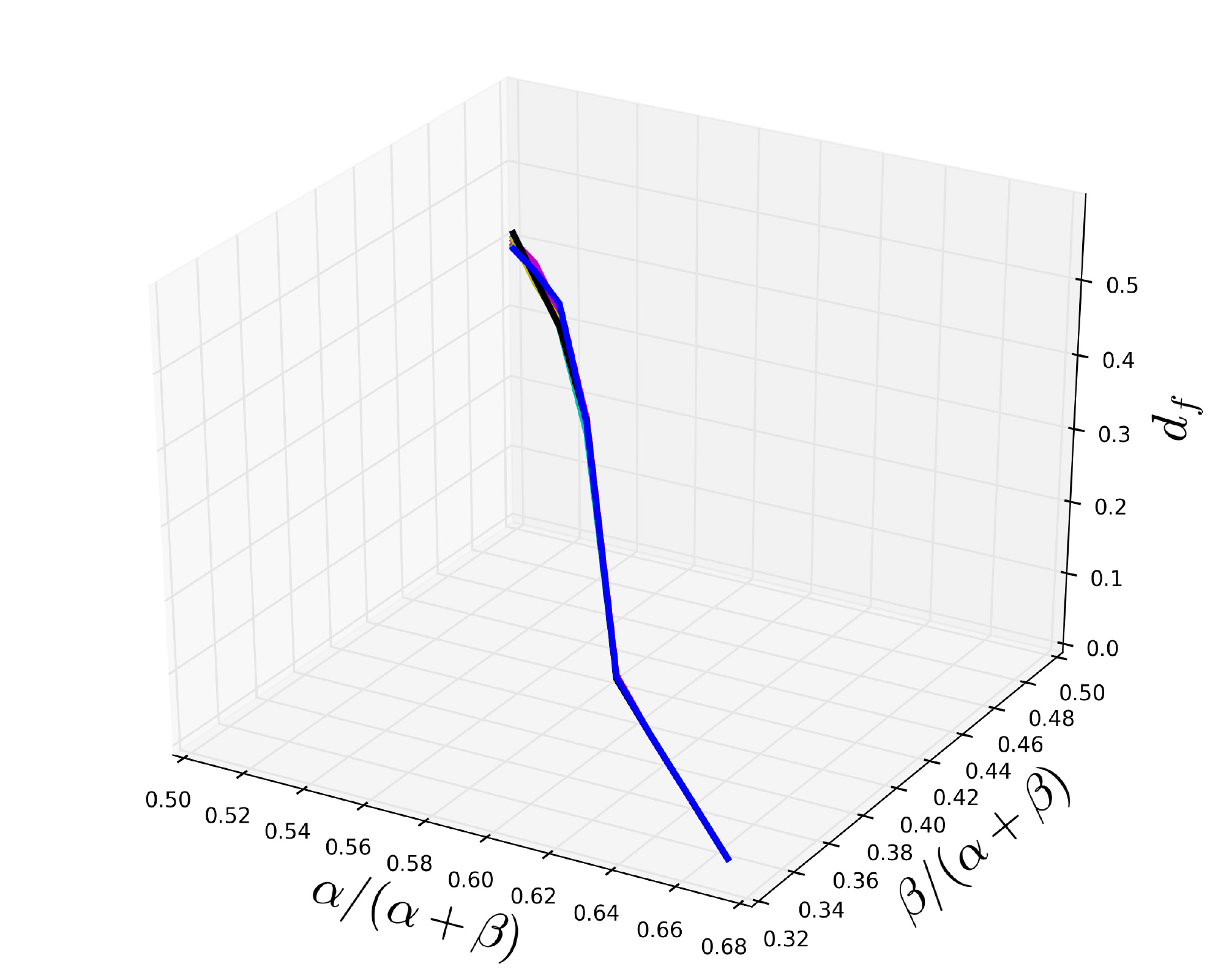}}
(d)\subfloat{\includegraphics[width=3in]{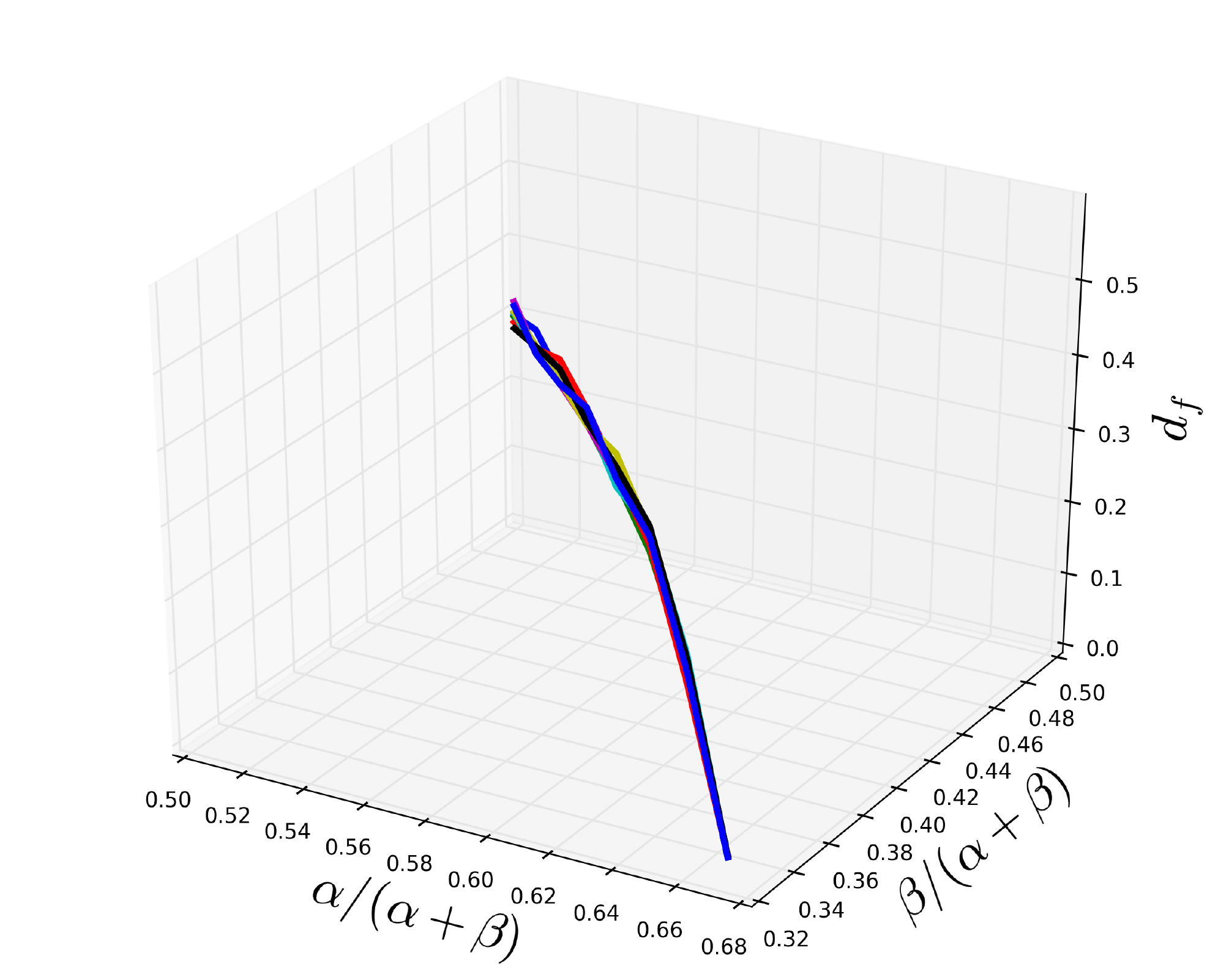}}
}
\caption{Final density of frustrated agents $d_f$ against reward ratio $\alpha/\beta$. (a) $\rho=0.6$ (b) $\rho=0.5$ (c) $\rho=0.4$ (d) $\rho=0.3$}
\end{figure}
As can be see from the four cases, the theoretical limits for $\rho$
to reach full satisfactory equilibria may not be quantitatively correct, but most importantly the analytical prediction demonstrate to be qualitatively correct and even reasonably accurate. 
Cases (a) and (c) are symmetric, demonstrating the symmetry of
the equilibria. In case (d) the full satisfactory configuration is
never reached, since the 1-preferences density is too low compared
to $\nicefrac{\beta}{\alpha+\beta}$, even if it is higher than $\nicefrac{\alpha}{\alpha+\beta}$.

\subsection{Anticoordination Game}

The case of AG under best response dynamics is peculiar because, as discussed in \cite{roca:2009b}, the fact that agents try to anticoordinate leads to unrealistic outcomes when the population is homogeneous. Let us keep in mind that
agents are given exclusively informations about the distribution of preferences but not about actions, so they
act to maximize their payoff expecting that neighbors are going to take their preferred action.
In the case of homogeneous distributions, for example
when the whole network is made of 1-preferences, every agent knows
that he has to anticoordinate with a neighborhood full of 1-preferences,
the result is that he will obviously choose action 0, but this happens
with every agent in the network. For heterogeneous distributions,
the more the connectivity of the graphs, the higher the reward ratio
has to be to allow hybrid equilibria to appear in the final configuration,
which means that connectivity fosters specialized equilibria, while
a large reward ratio, as usual, helps agents to keep satisfied and
not change their action. Therefore, anticoordination is reached easier when connectivity is low and reward ratio is high. These
conditions for anticoordination lead to strong outcome differences.
These same conditions allow satisfactory equilibria to appear. Of course the dynamics shown above for homogeneous
distributions give very large values for agents frustration, since
the loss of information about neighbors actions makes them totally
blind about what is going on. This implies that, while they try to
anticoordinate between same preference neighbors, they end up being
totally coordinated on the same undesired action. In this sense, it turns out that an equal distribution
of preferences is optimal to reach anticoordination, since agents think
that half of the neighborhood is like them so they are
not pushed to change their action to maximize their payoff.
For $\alpha/\beta=1$, connectivity does not affect at all the dynamics
of the network, but satisfaction is difficult to achieve because
agents are free to change their actions without changing
their payoff, and correspondingly there are some outcomes that show full satisfaction
when the distribution is close to 50-50, but by no means are all of them  satisfactory. 
On the contrary, when $\alpha/\beta=2$, frustration is avoided in the most
of the cases if the distribution is heterogeneous, and low connectivity
helps agent to avoid frustration because they can maintain their liked option.
Differently from the same experiment in the complete information
framework, here anticoordination is harder to achieve due to the loss
of information about the neighbors actions, but satisfactory equilibria
appear with some restrictions on reward ratio and connecitivity,
which did not appear with complete information.

\section{Conclusions}

In this paper, we have presented the results of a numerical simulation program addressing the issue of preference in network games from an evolutionary viewpoint. We have considered both coordination and anti-coordination games, as well as different network structures, including random and scale free graphs. We have also studied two dynamics, best response and proportional imitation,  which are more economic-like and biological-like, respectively, in order to assess the effects of noise and of a local perspective on decision making. Finally, we completed the picture by looking at two informationalcontexts, complete and incomplete. This program has allowed us to address the research questions we pointed out in the introduction. Thus, beginning in order of generality, regarding the question about the effects of preferences, a first, general finding is that in all scenarios the heterogeneous model behaves under evolutionary dynamics much closer
to the expectations from economic theory \cite{key-2,hernandez:2017} than the homogeneous one studied in \cite{us:2014}. Beyond this broad finding, it is important to point out that our model leads to a number of specific predictions which we summarize below. 

Let us now summarize our results about the cases of complements and substitutes. For the case of coordination games, we have observed that both types of dynamics lead to full coordination for
a wide range of compositions of the population. This is in contrast with the homogeneous case, in which the outcome of proportional imitation was always coordination in the risk-dominant, less benefitial action. Here agents tend to coordinate in the action that is preferred by the majority, which leads to a better payoff for the population as a whole, even if the minority is choosing the action they dislike.
When there are two preferences in the population, there are only mixed equilibria when the composition is approximately in the range 40\%--60\% of one type. In turn, this implies that equilibria are never satisfactory, in the sense that for any population composition there will always be frustrated agents playing the action they dislike. This problem aggravates in the already mentioned 40\%--60\% range, particularly for low $\alpha/\beta$ values; {a higher reward for the preferred action leads to players sticking to their preferences}, reducing the degree of coordination but at the same time lowering global frustration. 
Connectivity also plays a fundamental role in the achievement of coordination: Indeed, more connected networks result in full
coordination even in contexts of evenly split population, specially
when the reward ratio is kept small, i.e., when preferences are not particularly salient. In this respect, we observed that scale free networks with low degree are not connected enough to permit
the development of full coordination, and a higher density of ties between
individuals would be needed to let them achieve higher efficiency. 

Moving to anti-coordination we have observed that, also for both dynamics, the final states of the model are better in the sense that players do choose the opposite action as their partners. When interaction is of this type, particularly when the reward for choosing the preferred action is large, the amount of frustration is lower than that observed in the coordination problem. This is not what takes place when the reward is small: in that situation, players do anti-coordinate but the action they choose is determined by their surroundings more than by their own preferences, which in the end makes a large fraction of players unsatisfied. It is also interesting that connectivity, while still playing a role, has a less determinant influence on anticoordination than in coordination. As for the dynamics, when there is a large majority of one of the preferences in the population, we have observed that, somewhat counterintuitively, the whole population anticoordinates, as their local update do not really allow them to realize that they are in fact a majority. 

Finally, information is also very important to understand the effect of preference in strategic interactions on networks. When players have only information about the global composition of the population but not of their immediate partners, both coordination and anticoordination become more difficult, except in the extreme cases of a larger majority of one of the preferences or of an evenly split population. Because of different mechanisms we have discussed along the text, in wide population ranges there are very few frustrated players, and for large reward rations we have even observed many satisfactory hybrid equilibria, i.e., with no frustration whatsoever. Interestingly, we have also found that connectivity is beneficial in this case, as the actions players choose from their knowledge of the global fraction of preferences correspond better to a more populated neighborhood (thus mimicking the behavior of a mean-field model). 

In closing, we would like to note that our conclusions point to the soundness of the predictions made from standard economic theory and, therefore, to the applicability of the results we are presenting to real life situations. One particularly appealing conclusion is that, as the economic and biological dynamics yield similar results, our findings may have a much wider applicability that purely human societal issues and may be relevant, for instance, when different strains of a bacteria need to coordinate in producing some chemical. Focusing on the interactions between people, our results are particularly illuminating for the case of coordination, where we have seen that connectivity is benefitial for coordination. This indicates that in social situations where preference gives rise to conflict, one possible way to decrease the level of conflict and help people reach consensus is to increase the relations among both communities. Interestingly, recent experiments \cite{goyal:2017} show that when every player is connected with every other one, even when the population is close to a 50-50 composition full coordination is reached (but not always, some instances of hybrid equilbria have also been observed occasionally). This suggests that in fact the range in which we have found hybrid equilibria may vanish both in the very large size limit and when the network is fully connected. 
It is important to stress that, in the discussion of the results in \cite{us:2014}, up to four economic-style explanations were proposed, only to be discarded because they disagree in one way or another with the experimental results. Therefore, we are providing here a starting point for another approach that can be more fruitful, although its application to the results in \cite{us:2014} in full would require an extension to the case where subjects choose their own links. Similar experiments done on the networks we are studying here, which are amenable with similar laboratory setups, should shed light on the accuracy of our results and confirm or disprove the validity of the evolutionary approach to an economic-like problem. On the other hand, the downside of such a socially efficient outcome is a large minority taking an action they do not like (an issue that might not arise if what is wanted is anticoordination). In this respect, the only way to nudge the population to a better individual situation would be to decrease the saliency of preferences, by making more valuable the alternative choice. We hope that our study encourages more work both on the understanding of the effects of preference in a highly connected work and how to use them to achieve better societal outcomes both at the individual and at the global level. 

\ack

This work was partially supported by the EU through
FET-Proactive Project DOLFINS (contract no. 640772) and FET-Open Project IBSEN (contract no. 662725), and by grant FIS2015-64349-P (MINECO, Spain / FEDER, UE).

\section*{References}
\bibliography{thebibliography}

\end{document}